\documentclass[fleqn, usenatbib,onecolumn]{mnras}
\usepackage{hyperref}

\usepackage{newtxtext,newtxmath}

\usepackage[T1]{fontenc}
\usepackage{ae,aecompl}

\usepackage{graphicx}	
\usepackage{amsmath}	
\usepackage{amssymb}	
\usepackage{cleveref}

\title[Pulsar Identification Using AI ]{Pulsar Candidate Identification with Artificial Intelligence Techniques}

\author[P. Guo et al.]{
Ping Guo,$^{1}$\thanks{E-mail: pguo@bnu.edu.cn (PG); fqduan@bnu.edu.cn (FD); wangpei@nao.cas.cn (PW); bityaoyao@bit.edu.cn (YY); yinqian@bnu.edu.cn (QY); xxin@bit.edu.cn (XX)}
Fuqing Duan,$^{1}$
Pei Wang$^{3}$, 
Yao Yao$^{2}$, 
Qian Yin$^{1}$, 
Xin Xin$^{2}$,\\
Di Li$^{3}$,
Lei Qian$^{3}$,
Zhichen Pan$^{3}$,
Lei Zhang$^{3,4}$,
and Shen Wang$^{3,4}$
\\
$^{1}$Image Proc. and Patt. Recog. Lab, School of Systems Science, Beijing Normal University, Beijing 100875, China\\
$^{2}$School of Computer Science, Beijing Institute of Technology, Beijing 100081, China\\
$^{3}$CAS Key Laboratory of FAST, National Astronomical Observatories, Chinese Academy of Sciences, Beijing 100101, China\\
$^{4}$ University of Chinese Academy of Sciences, Beijing 100049, China\\
}

\date{Accepted XXX. Received YYY; in original form ZZZ}

\pubyear{2017}

\begin{document}
\label{firstpage}
\pagerange{\pageref{firstpage}--\pageref{lastpage}}
\maketitle

\begin{abstract}
Discovering pulsars is a significant and meaningful research topic in the field of radio astronomy. With the advent of astronomical instruments, the data volumes and data rates are growing exponentially. This fact necessitates a focus on the artificial intelligence (AI) technologies that can mine large astronomical datasets. Automatic pulsar candidate identification (APCI) can be considered as a task of determining the potential candidates for further investigation and eliminating the noises of the radio frequency interferences or other non-pulsar signals. As reported in the existing literature, AI techniques, especially convolutional neural network (CNN)-based techniques, have been adopted for the APCI. However, it is challenging to enhance the performance of CNN-based pulsar identification because an extremely limited number of real pulsar samples exist, which results in a crucial class imbalance problem. To address these problems, we proposed a framework that combines a deep convolution generative adversarial network (DCGAN) with a support vector machine (SVM). The DCGAN is used as a sample generation and feature learning model, and the SVM is adopted as the classifier for predicting the label of a candidate in the inference stage. The proposed framework is a novel technique, which not only can solve the class imbalance problem but also learn the discriminative feature representations of pulsar candidates instead of computing hand-crafted features in the preprocessing steps. The proposed method can enhance the accuracy of the APCI, and the computer experiments performed on two pulsar datasets verified the effectiveness and efficiency of the proposed method. 

\end{abstract}

\begin{keywords}
pulsars:general, methods:statistical, methods: data analysis, methods:numerical
\end{keywords}



\section{Introduction}

Pulsars are highly magnetized rotating neuron stars, which emit a beam of electromagnetic radiation. A pulsar can be compared to a laboratory under extreme physical conditions and can be used as a probe for a wide range of physics and astrophysics research, such as that of the equation of state of dense matter, properties of the interstellar medium, dark matter and dark energy, stellar evolution, and formation and evolution of binary and multiple-star systems. Especially, a spatial array of millisecond-period pulsars can be used as a gravitational-wave telescope that is sensitive to radiation at nanohertz frequencies. Discovering new pulsars using modern radio telescopes survey projects, such as the Parkes multi-beam pulsar survey (PMPS) \citep{manchester2001parkes}, high time resolution univers (HTRU) \citep{burke2011high},  and pulsar Arecibo lband feed array survey(PALFA) \citep{deneva2009arecibo} \textit{et al}., is an important and meaningful task in astronomical studies.
In particular,  the Five-hundred-meter Aperture Spherical Telescope (FAST) is expected to revolutionize pulsar astronomy \citep{nan2011five}, and the FAST 19-beams drift-scan pulsar survey is expected to discover $\sim$1500 new normal pulsars and about 200 millisecond pulsars (MSPs) \citep{smits2009pulsar}. The planned Commensal Radio Astronomy FAST Survey (CRAFTS) \citep{DiIEEE} is expected to discover many millisecond pulsars and  will thus make significant contribution to the PTA experiment. FAST has so far detected more than one hundred of promising pulsar candidates since the commissioning, 99 of which have been confirmed \footnote{\url{http://crafts.bao.ac.cn/}}.The first of pulsars discovered by FAST \citep{Qian2016firstPulsar} was PSR J1900-0134 from  observation data of Aug. 2017. Its pulse period is $1.8s$, and itsdispersion measure is $188 pc$ $cm^{-3}$; the first MSP discovery was made by FAST on Feb. 2018 coincident with the unassociated gamma-ray source 3FGL J0318.1+0252 from the FAST-Fermi LAT collaboration (Atel 11584). The two new discovered pulsars symbolize the dawn of a new era of systematic discoveries by Chinese radio telescopes.
The observation data of radiation signals from radio pulsar surveys are processed and folded into diagnostic plots, which are referred as pulsar candidates. Traditionally, human experts read these plots to determine if the corresponding candidates are from real pulsars or non-pulsar noises, and then perform further verifications on prospective candidates in hope of new discoveries.

Generally, one modern pulsar survey project can produce in total millions of candidates. Identifying these candidates by human eyes cannot meet the necessary for ranking the candidates produced by large surveys, and it is impossible to accomplish the further requirement of the real time data processing. Automatic scoring or artificial intelligence related methodologies have been successfully applied in the radio astronomy study, especially in the pulsar candidate selection. In the past years, 
\citet{lee2013peace} proposed a linear combination of six different quality factors to calculate a score for each candidate and rank the candidates according to the scores.
\citet{morello2014spinn} presented six hand-crafted features and used them to train a single-hidden-layer neural network for the binary classification of candidates.
\citet{zhu2014searching} proposed a pulsar image-based classification system (PICS), which takes the candidate plots as inputs to train single-hidden-layer artificial neural networks (ANN), histogram plots as the input to train support vector machines (SVMs), and two-dimensional images as the input to train convolutional neural networks (CNNs). The predictions of all the classifiers are then assembled together with a logistic regression classifier to construct the PICS.
\citet{Wang2019Plusar} proposed the use of a residual neural network to replace the simple 5-layer CNN architecture and achieved a high performance for the FAST drift-scan survey representations. Artificial intelligence techniques emulate human experts and can automatically learn the mapping relationships among the samples and labels from a training dataset labelled by human experts.

However, the performance of these methods is generally limited due to the data incompleteness or sample class imbalance. In fact, the number of positively labelled samples, i.e., real pulsar candidates, is severely insufficient. If a CNN architecture is similar to the LeNet-5 \citep{lecun1998gradient-based}, which contains two convolution layers, two max-pooling layers and a fully connected layer, owing to the very limited real pulsar data samples, such CNNs cannot be trained to learn satisfactory representations. 
In recent years, in the research field of artificial intelligence, the generative adversarial network (GANs) \citep{goodfellow2014generative} framework has become remarkably popular and proved to be effective for recovering the training data distribution.
When learning converges, the generative model (generator, \textit{G}) will generate new samples that  are very similar to those in training data.
Furthermore, the discriminative model (discriminator, \textit{D}) can predict the probability that an input sample comes from the training data.
Radford \textit{et al.} introduced deep convolutional generative adversarial networks (DCGANs) \citep{radford2015unsupervised} that adopt CNNs as the generator and discriminator. 

As we know, to fully train a deep CNN model such as AlexNet \citep{krizhevsky2012imagenet},a large number of training samples is required. Furthermore, if a training dataset with class imbalance is employed, a discriminator with prejudice will be obtained, which will recognize most unseen samples as a domain class member. Unfortunately, the pulsar candidate sample data is a class imbalanced dataset because the number of positive (true) pulsar candidate samples is extremely limited compared to the large number of non-pulsar samples. For example, the HTRU Medlat Training Data\citep{morello2014spinn}consists of precisely 1; 196 positive pulsar candidates and 89; 996 non-pulsar samples. The reason the number of real-pulsar candidates is considerably smaller than that of non-pulsar candidates is that real pulsars are difficult to discover. However, a modern pulsar survey could generate millions of false candidates in a short time. 

To combat such imbalanced training data, one of the tactics is to generate synthetic samples. In this work, we propose a framework named  DCGAN+L2\footnote{L2 stands for L2 norm.}-SVM to address  this crucial class imbalance problem, and the framework is illustrated in Fig.\ref{fig:figure1}.
In  our DCGAN+L2-SVM framework, a DCGAN model is trained with  all the positive pulsar samples and an equal number of non-pulsar samples.
The middle layer activation values of the discriminator in  learned DCGAN model are regarded as the deep features of the input samples.
A L2-SVM linear classifier is trained with these deep features of the positive and negative pulsar samples.  
In the inference stage,  the label of the input candidate is predicted using the trained L2-SVM linear classifier.
Here, the reason why we choose DCGANs are two folds.
On one hand, DCGANs  demonstrated reliable results for the natural image classification  task \citep{radford2015unsupervised}, in which they could extract a discriminative feature representation.
On the other hand, DCGANs demonstrate a satisfactory ability to converge in the training stage.

The following parts of this work are organized as follows:

In section~\ref{Sec:Automatic}, we  briefly introduce what is a pulsar candidate (see section~\ref{Sec:Automatic1}) and the task of automatic pulsar candidate selection (see section~\ref{Sec:Automatic2}), later provide a review of AI-based automatic selection methods (see section~\ref{Sec:Automatic3}).
In section~\ref{Sec:BasicModels}, thefundamental machine learning models including ANNs (see section~\ref{Sec:BasicModels1}), CNNs (see section~\ref{Sec:BasicModels2}), and GANs (see section~\ref{Sec:BasicModels3}) are briefly explained in short.
In section~\ref{Sec:DCGANAPS}, our proposed  DCGAN+L2-SVM model is introduced in detail,  including the definition of the  DCGAN (see section~\ref{Sec:DCGANAPS1}), specific network architecture of the proposed method (see also section~\ref{Sec:DCGANAPS1}),  and the proposed learning strategy (see section~\ref{Sec:DCGANAPS2}).
Next, the results of experiments performed on two pulsar datasets are processed and analysed  in ~\ref{Sec:Results}.
 We first introduce the two pulsar candidate datasets and evaluation metrics in section~\ref{Sec:Results1}.
Subsequently, we describe the details of the experiment settings including the DCGAN+L2-SVM parameters and data splits in section~\ref{Sec:Results2}.
In section~\ref{Sec:Results3}, we report and analyses evaluation results of DCGAN+L2-SVM  on two test datasets. This part demonstrates  the advantages of our proposed method in improving the pulsar selection objective evaluation metrics.
In section~\ref{Sec:Results4}, we develop a  specific Conditional-DCGAN method to proof the capability of DCGAN in generating usable synthetic samples.
In section~\ref{Sec:Results5}, we investigate the convergence of discriminative capability of DCGAN based feature representations during its training process.
In section~\ref{Sec:Results6}, more discussions and further work considerations are presented.
Section~\ref{Sec:Conclusion}presents the conclusions.

%
\section{Automatic Pulsar Candidate Selection}
\label{Sec:Automatic}

In this part, we first briefly introduce the pulsar candidate search task, and we later model the automatic pulsar candidate identification as a classification problem in the context of machine learning. Finally, we provide a review of the machine learning-based selection methods.

\subsection{Pulsar Candidates}
\label{Sec:Automatic1}
Modern radio telescope surveys are applied to search for the periodic broadband signals from the universe space that shows signs of dispersion. Analysing these signals provides a way to discover new pulsar stars in the universe, and each received signal is recorded and processed in the pipeline of telescope surveys.

As described in the existing literature, the datasets in which the pulsars are searched are time series of the total power per frequency channel, typically referred to as filter bank data. A typical pulsar searching pipeline used for the processing of the dataset consists of the following stages:

\begin{itemize}
\item The first step is to find and mark the radio frequency interference (RFI) in the data. The data is analysed in blocks of time/frequency to flag or replace the affected samples with averaged values for the persistent narrowband RFI and transient broadband RFI. After mitigating the RFI, the dispersion effect must be compensated (de-dispersion) for the frequency-dependent time delay induced by the process of propagation in the interstellar medium \citep{Lorimer2005}.
It should be noted that the RFI mitigation cannot completely remove the RFI. This aspect is a significant problem in pulsar searching and is the reason for the large number of non-pulsar candidates existing in the data. Consequently, our pulsar selection methods to classify candidates should successfully exclude the RFI generated candidates.

\item The second step is to process de-dispersion problem. When performing the searches, we do not know the dispersion measure (DM) of the pulsars; therefore, a wide range of trial DMs are corrected under an optimized de-dispersion plan, and a periodic signal search of each de-dispersed time series is performed.

\item This step is known as a periodicity search.  The most common is to perform an Fourier analysis \citep{lyon2016fifty} and then incoherently summarize possible harmonics in fundamentals to increase the signal to noise ratio, i.e. the original power spectra and the composite spectra formed by summing harmonics are each searched for isolated periodicities \citep{Cordes2006}. But for the binary pulsars, the period will change due to the Doppler shift from orbital motion, the correlation method or acceleration search technique  are often used to mitigate pulse smearing across multiple Fourier bins. After each of the de-dispersed time series have been searched, the resulting candidate signals must be sifted through in order to determine the likely pulsar candidates and remove ones which are unlikely to be real. Typical tactics used are to remove candidates which have periods of known RFI, do not appear at contiguous trial DMs, or are detected at only a single DM value. 

\item  In this step, the candidates that make it through the sifting process are inserted into a candidate list and diagnostic plots, which are stored for further analysis. The basic candidate consists of a small collection of characteristic variables, including the S/N, DM, period, pulse width, and averaged pulse profile, which are used to describe how the signal persists and distributes throughout the time and frequency domains.

\end{itemize}

The above mentioned pulsar search steps are usually implemented  via a pipeline software, such as PulsaR Exploration and Search Toolkit (PRESTO)\footnote{http://www.cv.nrao.edu/~sransom/presto/}.
PRESTO is a pulsar search and analysis software which has helped discover more than 600 pulsars, including over 200 recycled and binary pulsars.

\begin{figure*}
	\begin{center}
		\includegraphics[width=0.8\linewidth]{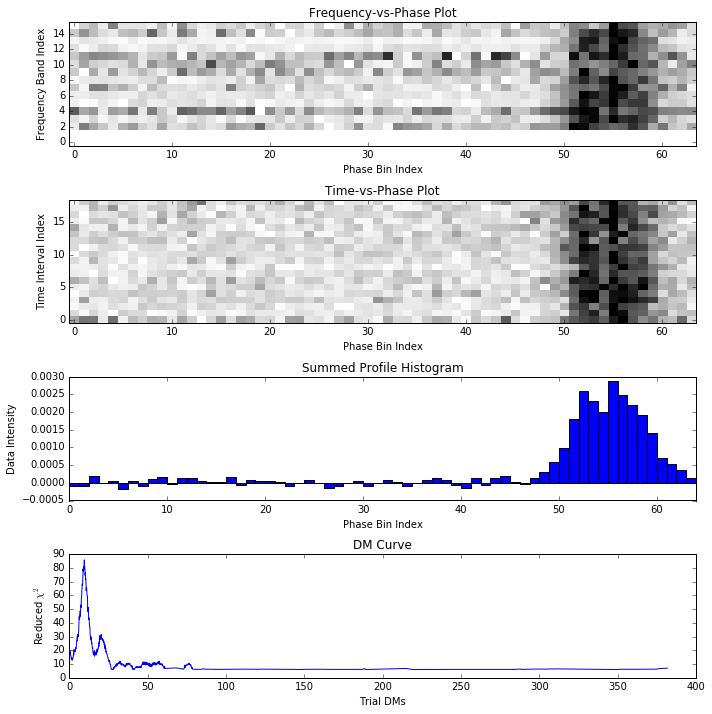}
	\end{center}
	\caption{An example of positive (true) candidate from HTRU medlat dataset that is folded from signals of a real pulsar. For a real pulsar, there is a bold vertical stripe on the right side of both frequency-vs-phase and time-vs-phase plots. There is one peak in the summed profile histogram. And DM curve peaks at a non-zero value.}
	\label{fig:figure2}
\end{figure*}

A folded pulsar candidate is a group of diagnostic plots, which are represented by a two-dimensional matrix. These diagnostic plots can be considered as images to be processed using the classification models. The pulsar candidate diagnostic representations mainly include the \textit{summed profile histogram} (SPH), \textit{time-vs-phase plot} (TPP), \textit{frequency-vs-phase plot} (FPP) and \textit{dispersion-measure (DM) curve}.
\begin{itemize}
	\item \textit{Summed profile histogram} is an intensity-vs-phase pulse profile. This histogram is obtained by adding the data over both time intervals and frequencies. The profile of a real pulsar usually contains one or more narrow peaks. Pulse phase is how far we are through the pulsar's rotation. Pulse phase equals 0.5 means that the pulsar has gone through half a rotation, and 1.0 means one full rotation.
	\item \textit{Frequency-vs-phase plot} is obtained by summing the data over time intervals. For a real pulsar,one or more vertical lines should be present,  which indicates a broadband signal was observed. If the frequency-vs-phase plot is summed over the frequency bands, we can obtain the summed profile histogram.
	\item \textit{Time-vs-phase plot} is  obtained by summing the data over the frequencies. For a real pulsar, one or more vertical stripes should be present, which indicates that a pulsed signal was observed.  When summing the time-vs-phase plot over the time intervals, the summed profile histogram can be obtained.
	\item \textit{DM curve} is a histogram of the trial DMs against the corresponding $\chi^2$ values. For a real pulsar, DM curve peaks at a nonzero value.
\end{itemize}

\begin{figure*}
	\begin{center}
		\includegraphics[width=0.8\linewidth]{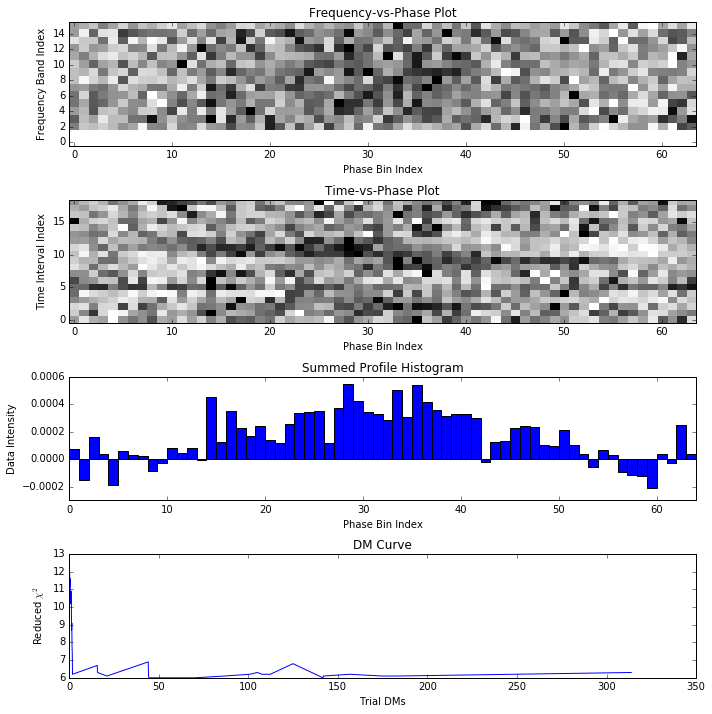}
	\end{center}
	\caption{An example of negative (false) candidate from HTRU medlat dataset, which is folded from signals of a non-pulsar. For a non-pulsar, there is no obvious vertical stripe in neither frequency-vs-phase nor time-vs-phase plots. There are also no notable peaks in the summed profile histogram. And DM curve peaks at zero value.}
	\label{fig:figure3}
\end{figure*}

Fig.\ref{fig:figure2} shows an example of positive (true) pulsar candidate from HTRU medlat dataset, which comes from a real pulsar.
Fig.\ref{fig:figure3} illustrates an example of negative (false)  candidate from HTRU medlat dataset, which is folded from a non-pulsar signal.
From these two figures, we can see that positive (true) and negative (false) candidates have different characteristics in diagnostic plots.

\subsection{Modelling the Automatic Selection of Pulsar Candidates}
\label{Sec:Automatic2}
Pulsar candidate selection is the task of determining the prospective candidates for further verification and excluding the noise candidates, which might be radio frequency interferences or other non-pulsar signals. For most candidates, the diagnostic plots show notable difference between a real pulsar and a non-pulsar, for example Fig.\ref{fig:figure2} is considerably different compared to Fig.\ref{fig:figure3} according to human.
However,  some exceptions exist; for example, when a real pulsar points at the Earth throughout most of its rotation, the intensity may spread out in plots. This phenomenon makes it more difficult to determine if a candidate is real. As another example, some pulsars may have not only one peak in plots but also many peaks at different phases, which makes it considerably more difficult to classify candidates. This difficulty leads to the several false positive candidates not being excluded using the cutting-off or sifting methods. In addition, modern radio observatories produce millions of candidates in a short time. To this end, AI techniques have demonstrated excellent capability in the image classification tasks. Therefore, the automatic selection of pulsar candidates by using AI techniques is a meaningful and important method for discovering new pulsars.

The automatic pulsar candidate selection task can be modelled as follows: A pulsar candidate for training  is denoted as a tuple of five elements $x=(x_{subints}, x_{subbands}, x_{profile}, x_{dm}, y)$, where $x_{subints}$ refers to a time-vs-phase plot, $x_{subbands}$ refers to a frequency-vs-phase plot, $x_{profile}$ for each observation is formulated by summing all the frequency channels and subintegrations and adding the highest signal-to-noise ratio (S/N) profiles to form the standard profiles for each observation frequency; $x_{dm}$ is the corresponding DM curve; and  $y$ is the ground-truth label of $x$, that is,a real pulsar or non-pulsar.

A prediction function $f$ is learned based on the training data, which can be expressed as in equation~\ref{eq:eq1}.
The inference error of the prediction function $f$ is $L(f(x)-y)$.
For all samples in training dataset, the average inference error should be minimized.
The optimal solution of the prediction function is $f_*$.

\begin{equation}
	\label{eq:eq1}
	f_* = argmin_f \frac{1}{n} \sum_{x \in X} L(f(x)-y),
\end{equation}
where $x$ is a pulsar candidate sample in training dataset $X$,
$y$ is the corresponding label of that candidate,
$n$ is the number of training samples in $X$,
and $L$ is a loss function defined in the training algorithm, and $f_*$ is a well-trained optimal prediction function.
In the inference stage, $y_{predict}=f_*(x_{test})$ makes the real pulsar (true) or non-pulsar (false) label prediction for an unknown input candidate $x_{test}$.

\subsection{Automatic Pulsar Candidate Selection with Machine Learning Techniques}
\label{Sec:Automatic3}
Recently, scoring and machine learning-based methods have demonstrated excellent  capabilities in the automatic pulsar candidate selection task. 
\citet{lee2013peace} proposed a scoring method called Pulsar Evaluation Algorithm for Candidate Extraction (PEACE), which computes a score using six different quality factors.
Their ranking method helped to find 47 new pulsars.
\citet{eatough2010selection} hand-crafted 12 features and trained a feed-forward single-hidden-layer artificial neural network.
This work helped to find one new pulsar in PMPS data.
\citet{bates2012high} adopted the 12 features from \citet{eatough2010selection} and 10 more from \citep{keith2009discovery} and trained a feed-forward single-hidden-layer artificial neural network with these 22 features.
Using this approach, 75 new pulsars were discovered.
\citet{morello2014spinn} proposed the Straightforward Pulsar Identification using Neural Networks (SPINN), in which they designed six features and used these features to train a feed-forward single-hidden-layer  neural network for candidate binary classification.
Their method contributed to 4 new pulsar discoveries. 
\citet{thornton2013high} also used a single-hidden-layer network as a classifier and the hand-crafted 22 features based on candidate plots and observation parameters were taken as the training input.
\citet{lyon2016fifty} designed 8 features from diagnostic histograms to describe pulsar candidate and proposed a fast decision tree model for online operation.
Their model helped to find 20 new pulsars.
In previous machine learning based method, artificial neural networks, especially single-hidden-layer networks played a key role.
\citet{wang2017new, guo2004pseudoinverse} proposed a pseudoinverse incremental representation learning algorithm for training neural networks and applied it to perform spectra pattern recognition. Their method could also be applied to pulsar search task.
\citet{zhu2014searching} developed a pulsar image-based classification system, in  which the candidate's histograms and plots  were taken as the input and three models were trained. These models included a single-hidden-layer network, an SVM and a CNN, and finally, all these classifiers were combined together with the logistic regression function to form an ensemble net. Among these methods, the CNNs demonstrate a satisfactory classification performance on two-dimensional diagnostic plots. Compared to other methods, the CNNs take the plots as the input and realize end-to-end learning; in contrast, the other methods require hand-crafted features for pulsar candidate selection. Furthermore, CNN based models performed better than other methods in prediction tasks  \citep{zhu2014searching}. 
In addition, by assembling all the classifiers, the accuracy can be improved over that corresponding to the use of a single classifier.

Although a CNN based model is powerful when using two-dimensional image data, training a deep CNN model requires more labelled samples. As we know, in a real scenario, labelled data is difficult and expensive to obtain. In the pulsar candidate selection task, the positive samples are limited because the number of discovered real pulsars are small and millions of candidates are negative samples. This fact leads to the occurrence of a crucial class imbalance problem. For machine learning-based techniques, insufficient training data cannot realize successful prediction function learning. Imbalanced samples make the prediction function biased for the class that has more training samples. Therefore, learning a discriminative classification model with a limited number of training candidates and class imbalance samples is an important and challenging issue.

\begin{figure*}
	\begin{center}
		\includegraphics[width=0.75\linewidth]{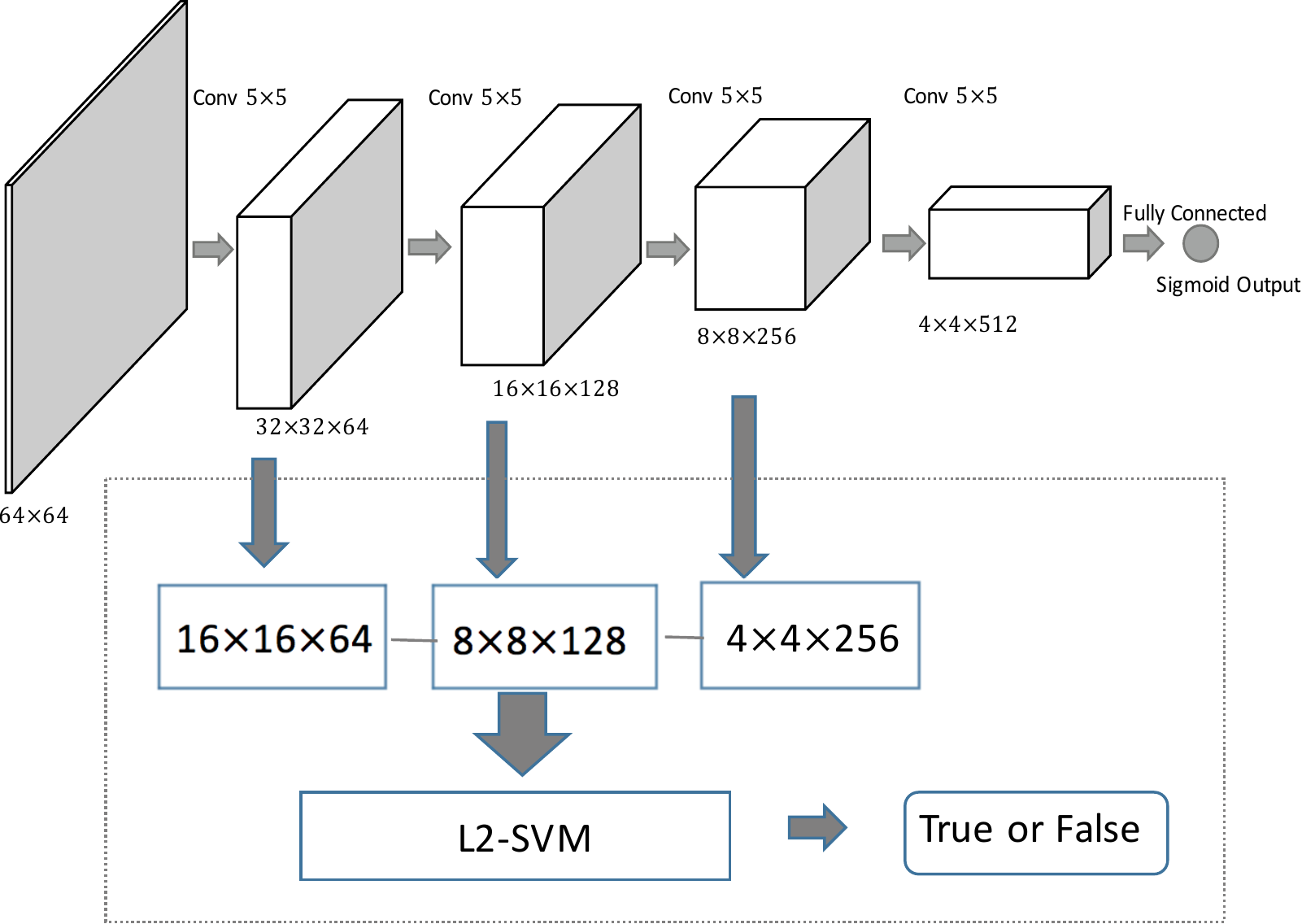}
	\end{center}
	\caption{An illustration of our DCGAN+L2-SVM method. The candidate's  hierarchical features are mapped with the discriminator of the trained DCGAN model, which consist of a 16-by-16-by-64 tensor, a 8-by-8-by-128 tensor, and a 4-by-4-by-256 tensor. A L2-SVM classifier is trained with  feature vectors chained three tensors. In testing stage, the label of input candidate is predicted with this trained L2-SVM classifier.}
	\label{fig:figure1}
\end{figure*}

\section{Basic Machine Learning Models}
\label{Sec:BasicModels}
In this section, several related basic machine learning models, including ANN, CNN, and GAN are briefly reviewed.

\subsection{Feed-forward Artificial Neural Networks}
\label{Sec:BasicModels1}
\begin{figure}
	\begin{center}
		\includegraphics[width=0.8\linewidth]{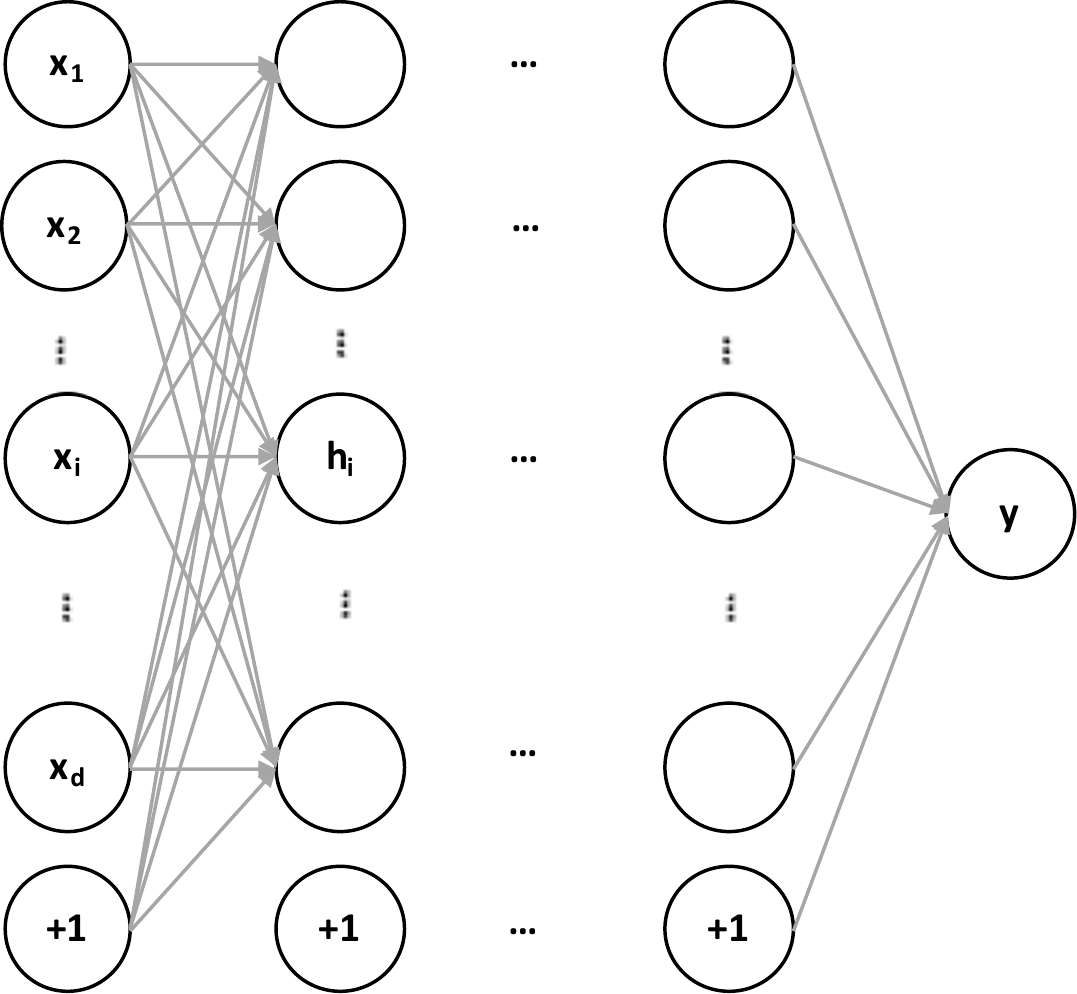}
	\end{center}
	\caption{An example of feed-forward fully connected artificial neural network.}
	\label{fig:figureFNN}
\end{figure}

A typical feed-forward fully connected artificial neural network consists of an input layer, several  hidden layers and an output layer.
An example architecture of this type of ANN is shown in Fig.~\ref{fig:figureFNN}.
The input layer takes as an input vector $\boldsymbol{x} \in \mathbb{R}^d$.
The elements in the former layer are combined with a weight matrix $\boldsymbol{w}_i \in \mathbb{R}^d$ and a bias $b_i \in \mathbb{R}$.
In Fig.~\ref{fig:figureFNN}, taking the $i^{th}$ neuron as an example, the output to the $i^{th}$ neuron in the second layer is
\begin{equation}
\label{eq:eqhi}
h_i = \boldsymbol{w}^T_i\boldsymbol{x} + b_i.
\end{equation}
Next the value $h_i$ is processed using an activation function, such as the sigmoid function, which can be defined as $g(h_i) = 1 / {(1+e^{-h_i})}$.
introduces non-linearity into the entire  ANN mapping, and  makes the output value lie in some specific range, for example, the  sigmoid function is monotonically increasing and its output value is in the range $[0,1]$.
After all neuron values in this layer are computed in this way, 
they are then fed into the next layer as an input vector.
A similar calculation processing is performed layer by layer, 
and the final value is computed as $y$ in the  output layer. 

In training stage, the goal is to learn all the weights and biases in all layers, 
and the classical error back propagation (BP) learning algorithm \citep{rumelhart1988learning}  is often adopted to train the  ``shallow'' ANNs.
Here shallow means that the networks only consist of a small number of hidden layers (one or two hidden layers).
When the number of  hidden layers becomes larger, usually  more than 5 layers, the ANN model will be considered as a deep neural network (DNN).
The deep fully connected network contains excessively  many parameters to be learnt, 
and the gradient in the back propagation algorithm may vanish in  deep architecture.
Some learning algorithms such as the  auto-encoders \citep{bourlard1988auto,bengio2009learning, guo2004pseudoinverse,wang2017new, Guo2017pilae} can be adopted  to train a deep neural network under the strategy of greedy layer-wise learning.
Since single  hidden layers ANNs have been extensively  studied  in previous automatic pulsar candidate selection works, \citet{eatough2010selection,bates2012high,thornton2013high,zhu2014searching,morello2014spinn},all adopted this type of  feed-forward ANN as the classifier model.

\subsection{Convolutional Neural Networks}
\label{Sec:BasicModels2}

	A CNN is a special type of feed-forward ANN. Unlike the feed-forward ANN, in which the output of the former layer is a linear combination of all the input elements  (presented in equation~\ref{eq:eqhi}), the CNN calculates the output only in a two-dimensional spatial neighbourhood, which is known as receptive field. Furthermore, an ANN does not share the weight matrices in one layer; however, the weight matrices in a CNN layer are shared. The shared weight matrices are called filters. The CNN takes the advantage of the local connectivity in the receptive field of two-dimensional images, and deep CNN architectures have been widely adopted in the research field of computer vision. Because a CNN based neural network architecture has the advantage of being able to handle images, it can be adopted as a better feature learning model for automatic pulsar candidate selection when the diagnostic plots are considered as images. In a previous work, \citet{zhu2014searching} demonstrated the capability and advantages of CNN for pulsar candidate selection task.

The CNN networks can be used as classifiers that can automatically classify pulsar candidates into different classes when these CNNs have been trained on training data. \citet{LeCun1989ZipCode} trained CNNs to recognize handwritten zip code, and \citet{LeCun1990Handwrite} trained CNNs to recognize handwritten digits.
\citet{krizhevsky2012imagenet} developed an 8-layer CNN to classify natural images from the ImageNet dataset \citep{imagenet_cvpr09}. This CNN  demonstrate excellent  capability in image recognition and  led to  the popularity of CNN in recent years.

\begin{figure}
	\begin{center}
		\includegraphics[width=0.8\linewidth]{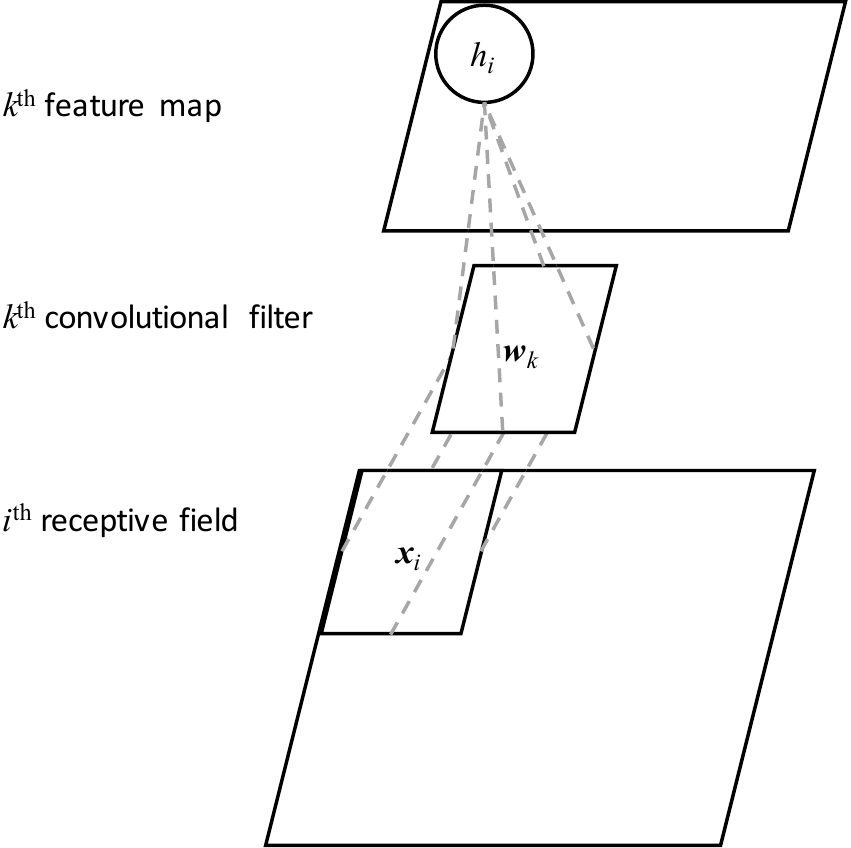}
	\end{center}
	\caption{An illustration of stride convolutional layer. $\boldsymbol{x}_i$ is an image block called  $i^{th}$ receptive field on a sliding window of input image. $\boldsymbol{w}_k$ is the $k^{th}$ convolution filter.  $\boldsymbol{x}_i$ and $\boldsymbol{w}_k$ have the same size. $h_i$ is the convolution output value corresponding to $\boldsymbol{x}_i$.}
	\label{fig:figureCNNill}
\end{figure}

Fig.~\ref{fig:figureCNNill} illustrates the working mechanism of a  CNN layer.
Denote the $i^{th}$ block (\textit{i.e.} receptive field) of input two-dimensional image as $\boldsymbol{x}_i$,
the $k^{th}$ filter as $\boldsymbol{w}_k$.
The output value $h_i$ of convolution operation corresponding to $\boldsymbol{x}_i$ is
\begin{equation}
\label{equ:convolution}
h_i = \sigma(\boldsymbol{w}^T_k \boldsymbol{x}_i + b_k),
\end{equation}
where $\boldsymbol{w}_k$ and $\boldsymbol{x}_i$ are both flattened into vectors, $b_k$ is a bias and $\sigma$ is an activation function, typically $\sigma(t)=max(0,t)$.
In \citet{zhu2014searching}, a hyperbolic tangent (Tanh) activation function is adopted.
The output image of a convolutional layer is called a feature map.
In Fig.~\ref{fig:figureCNNill}, the $k^{th}$ feature map is obtained by applying the $k^{th}$ filter on every block of the input image divided by a sliding window strategy.
Obviously, the size of the input image, stride of the sliding window and convolutional filter size all together determine the size of the output feature map.

\begin{figure}
	\begin{center}
		\includegraphics[width=0.8\linewidth]{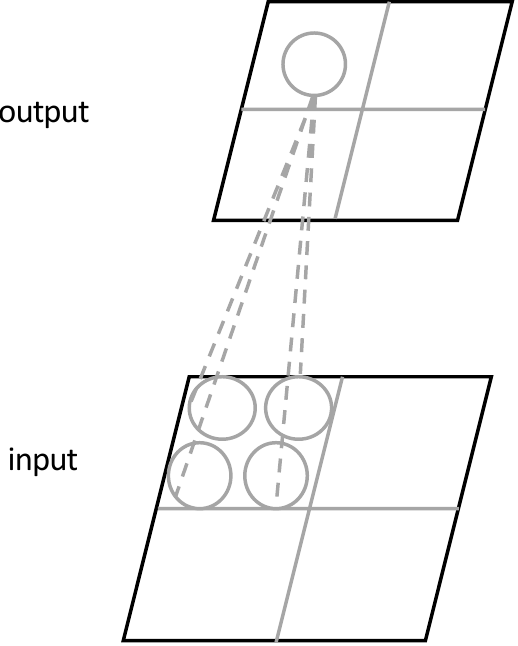}
	\end{center}
	\caption{An illustration of max pooling layer. Four values in a neighborhood are pooled into one value in the output.}
	\label{fig:figureMaxpool}
\end{figure}

After a convolutional layer, there usually is a pooling layer.
A pooling layer down-samples the feature maps into a smaller map and introduces certain invariant characteristics. A pooling layer is illustrated in  Fig.~\ref{fig:figureMaxpool}, where  four values in the input feature map are pooled into one value in the output.
Typically, the pooling method is a maximum or averaging function.
In \citet{zhu2014searching}, the max pooling is adopted, and 
the CNN architecture is a LeNet-5 - like network \citep{lecun1998gradient-based}:
The first and third layers are convolutional layers, the second and fourth layers are max pooling layers, and the last layer is a fully connected layer with a sigmoid output.


\subsection{Generative Adversarial Networks}
\label{Sec:BasicModels3}
Generative Adversarial Network (GAN) is a new framework proposed by Goodfellow et al. \citep{goodfellow2014generative, goodfellow2016nips}.
GAN originally aims to estimate generative models via an adversarial learning process.
In the process, two models are learned: a generative model \textit{G} called generator, which captures the data distribution of training samples, and a discriminative model \textit{D} called discriminator, which predicts the probability of an input sample coming from the real training samples rather than the fake data generated by \textit{G}.
For generator \textit{G} the training is to maximize the probability of discriminator \textit{D} making mistakes.
For discriminator \textit{D} the training is to minimize the probability of making mistakes.

The whole training is proceeded by fixing \textit{D} updating \textit{G}, and then fixing \textit{G} updating \textit{D} iteratively, and finally the learning  converges to a stationary status.

\begin{figure}
	\begin{center}
		\includegraphics[width=0.95\linewidth]{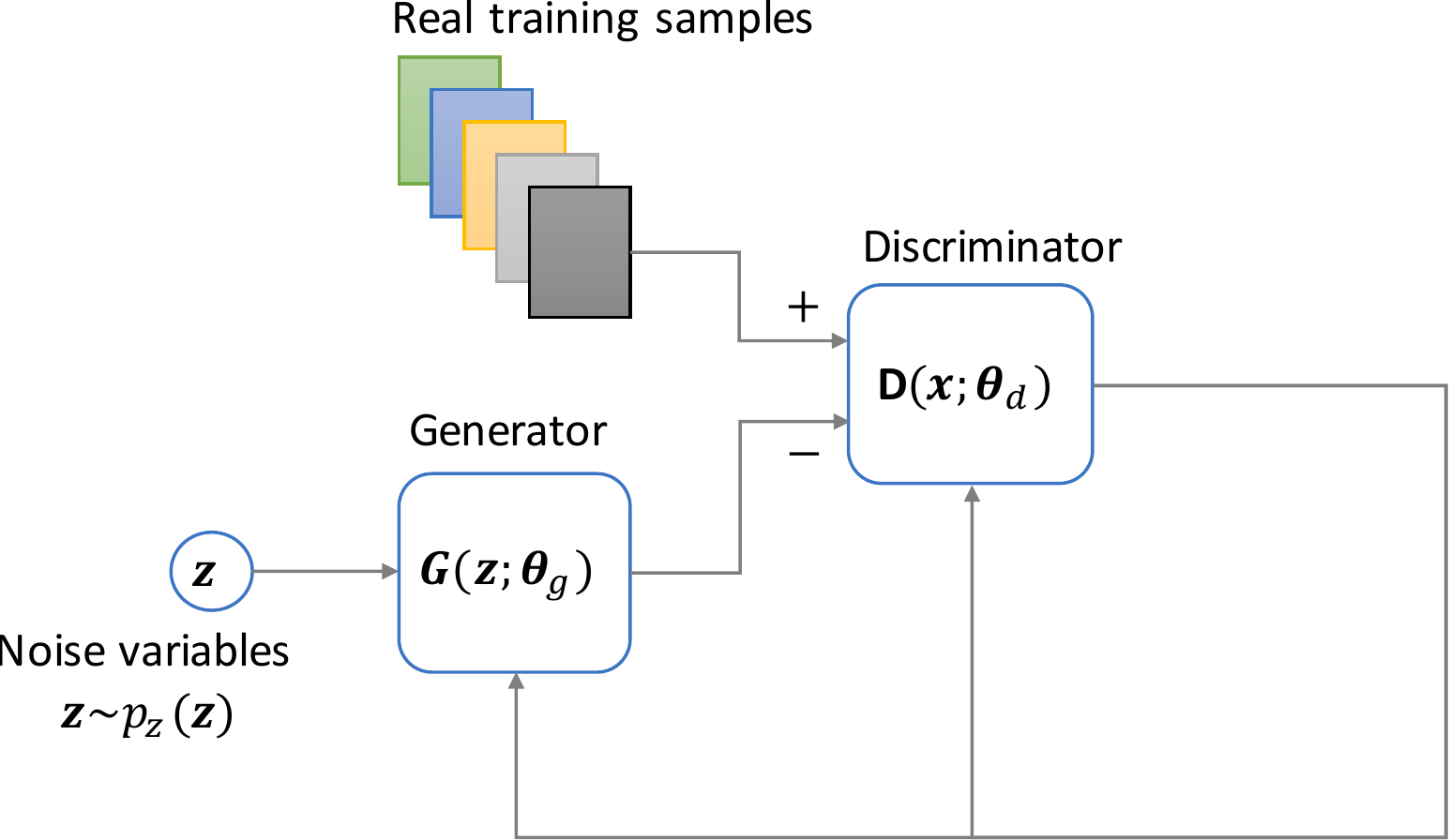}
	\end{center}
	\caption{Generative adversarial network framework. Noise prior distribution  $\boldsymbol{z} \thicksim p_z(\boldsymbol{z})$. $\boldsymbol{G}$ is the generator. $\boldsymbol{D}$ is the discriminator. Fix \textit{G}, \textit{D} is trained with real data samples and synthesized fake data by minimizing the probability of taking fake data as real ones. Fix \textit{D}, \textit{G} is trained with some synthesized fake data by maximizing the probability of taking fake data as real ones. }
	\label{fig:figureGANFramework}
\end{figure}

This procedure is illustrated in Fig.~\ref{fig:figureGANFramework}.
The noise variable $\boldsymbol{z}$ has a prior distribution $p_z(\boldsymbol{z})$, 
$\boldsymbol{z}$ is then mapped into the training data space with the generator $G(\boldsymbol{z};\theta_g)$.
The generated fake data samples and real training data are used together to train the discriminator $D(\boldsymbol{x};\theta_d)$.
The training of \textit{G} and \textit{D} are processed alternatively.

The process aims at optimizing the following value function $V(D, G)$: When fixing G, D is search, which maximize $V(D, G)$ and iteratively, when fixing D, G is searched,  which minimize $V(D, G)$,
\begin{align}
\label{equ:GANvalue}
min_{G}\;max_{D}\;& V(D, G) =  
\mathbb{E}_{x \thicksim p_{t}(x)} [log D(x) ]  \nonumber \\
& + \mathbb{E}_{z \thicksim p_{z}(z)} [log (1 - D( G(z))) ].
\end{align}
Where $p_{t}(x)$ refers to the training data distribution, and $\mathbb{E}[\cdot]$ denotes the expectation value which  is an averaged value on a mini-batch in practice.

As described by \citet{goodfellow2014generative, goodfellow2016nips}, the learning algorithm for the GAN framework can be summarized as follows:
\begin{enumerate}
\item[\textbf{A})] Fix generator \textit{G}, train discriminator \textit{D}, repeat $k$ times: 
	\begin{enumerate}
	\item[1)] from noise prior $p_z(\boldsymbol{z})$, sample a mini-batch of $m$ noise $\{\boldsymbol{z}^{(1)} \cdots \boldsymbol{z}^{(m)}\}$;
	\item[2)] from real training data, sample a mini-batch of $m$ samples $\{\boldsymbol{x}^{(1)} \cdots \boldsymbol{x}^{(m)}\}$;
	\item[3)] update discriminator parameter $\boldsymbol{\theta}_d$ by ascending its stochastic gradient:
	\begin{equation}
		\label{equ:ascending}
		\nabla_{\boldsymbol{\theta}_d}\frac{1}{m}\sum_{i=1}^{m}[log \textit{D}(\boldsymbol{x}^{(i)} )+log(1-\textit{D}(\textit{G}(\boldsymbol{z}^{(i)})))]
	\end{equation}
	\end{enumerate}
\item[\textbf{B})]  From noise prior $p_z(\boldsymbol{z})$, sample a mini-batch of $m$ noise $\{\boldsymbol{z}^{(1)} \cdots \boldsymbol{z}^{(m)}\}$;
\item[\textbf{C})] Fix discriminator \textit{D}, update generator parameter $\boldsymbol{\theta}_g$ by descending its stochastic gradient:
	\begin{equation}
	\label{equ:descending}
	\nabla_{\boldsymbol{\theta}_g}\frac{1}{m}\sum_{i=1}^{m}[log(1-\textit{D}(\textit{G}(\boldsymbol{z}^{(i)}))).
	\end{equation}
\end{enumerate}

The steps from \textbf{A}) to \textbf{C}) are updated iteratively until the maximum number of training iterations is reached.

Here the noise prior in DCGAN generator is different from the signal noise in pulsar candidates or RFI.
The pulsar signal noise does not cause any false candidates. Most false candidates are caused by the  RFI, which is different from the signal noise. The RFI is produced by a real signal , however, this signal is  not a pulsar signal.
The noise prior in the DCGAN generator  is used as a seed to generate  real pulsar samples. This noise prior add a little perturbation between different generated real pulsar samples.

\section{DCGAN-based Automatic Pulsar Candidate Selection}
\label{Sec:DCGANAPS}
This section describes the proposed DCGAN-based model shown in Fig~\ref{fig:figure1}, DCGAN+L2-SVM, including generator and discriminator neural network architectures.
\subsection{DCGAN+L2-SVM}
\label{Sec:DCGANAPS1}

DCGAN defines a family of GAN architectures that can perform stable training and to learn satisfactory  image feature representations in an unsupervised learning manner. 
In this work, we   take the time-vs-phase and frequency-vs-phase two-dimensional plots as the input of DCGAN+L2-SVM, and a L2-SVM linear classifier is trained  for predicting labels of the new input samples.  An illustration of the framework of DCGAN+L2-SVM is shown in Fig.\ref{fig:figure1}.


The architecture of generator and discriminator in DCGAN+L2-SVM follows the guidelines for the stable Deep Convolution GANs in \citet{radford2015unsupervised},
which includes the following items:
\begin{itemize}
	\item The stride convolutions \citep{springenberg2014striving} are used in discriminator to learn  spatial downsampling. The fractional-stride convolutions are used in generator to learn upsampling. No fully-connected or pooling are adopted in this step.
	\item The Batch Normalization \citep{ioffe2015batch} technique is used in  both discriminator and generator.
	\item  The ReLU activation function is used \citep{nair2010rectified} in all the layers of the generator except the  output layer, which uses the Tanh activation function.
	\item LeakyReLU activation function  \citep{xu2015empirical,maas2013rectifier} is used in all the layers of discriminator.	 
\end{itemize}

\begin{figure*}
	\begin{center}
		\includegraphics[width=0.9\linewidth]{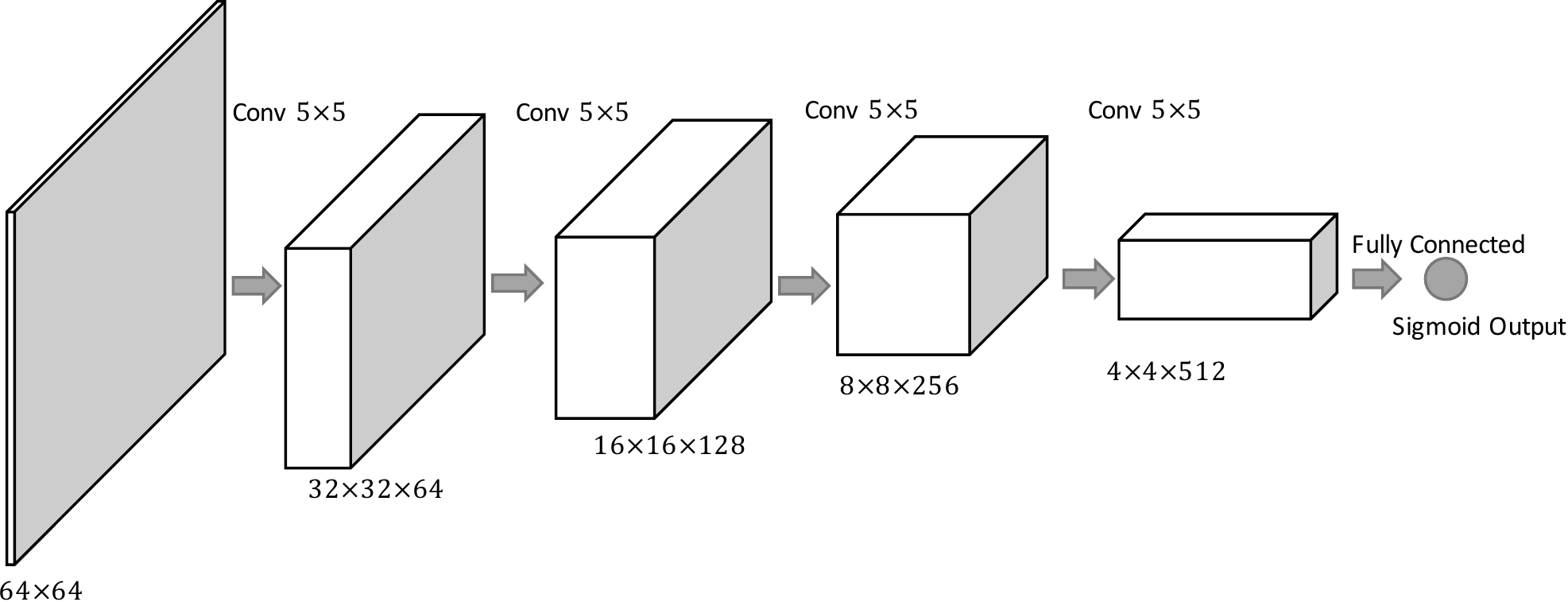}
	\end{center}
	\caption{DCGAN discriminator, $D$, trained in DCGAN+L2-SVM. The input samples are 64-by-64 gray images. There are four stride convolution layers. The final 4-by-4-by-512 features are fed into a single sigmoid output, which provides the probability of input image coming from real dataset.}
	\label{fig:figure4}
\end{figure*}
\begin{figure*}
	\begin{center}
		\includegraphics[width=0.9\linewidth]{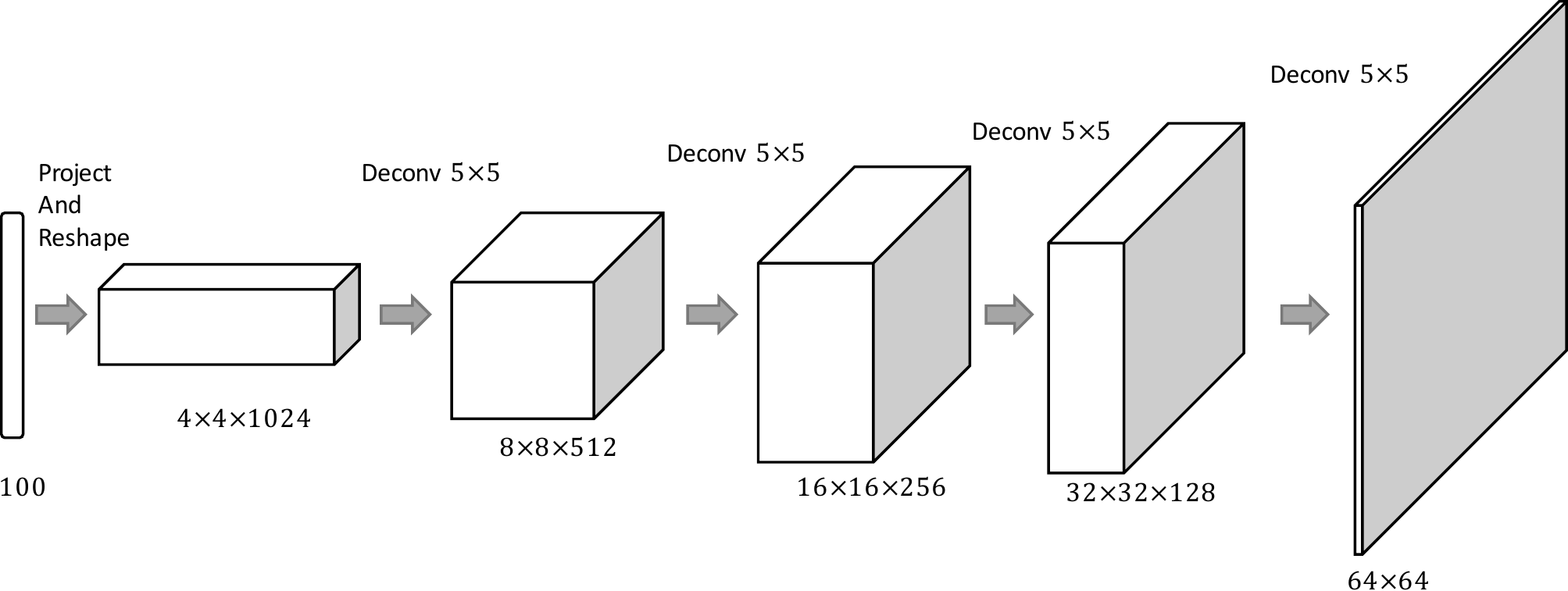}
	\end{center}
	\caption{DCGAN generator, $G$, trained in DCGAN+L2-SVM. The input is a 100 dimensional uniform noise distribution Z. Input is randomly projected and then reshaped into a 4-by-4-by-1024 tensor. There are four fractional-stride convolution layers (or deconvolution layers). The output, $G(z)$, is a generated 64-by-64 pixel gray image. }
	\label{fig:figure5}
\end{figure*}

Using our experience in artificial intelligence research, we design  the architecture of  both  discriminator and the generator as the one of four layers of convolution or deconvolution.
Specifically, the structure of the discriminator, whose input gray image size is 64-by-64, is shown in Fig. \ref{fig:figure4}.
The discriminator uses a serial of four stride convolutional layers and predicts the probability with a single sigmoid neuron output.
In the first convolutional layer, 64 kernels of size $5\times5$ are adopted, the stride is $2$, and the output feature map is a $32\times32\times64$ tensor.
In the second convolutional layer, 2 kernels of size $5\times5$ are adopted and the stride is $2$, this configuration results in  the output feature map becoming a $16\times16\times128$ tensor.
In the third convolutional layer, 2 kernels of size $5\times5$ are adopted , the stride is $2$ and the output feature map is a $8\times8\times256$ tensor.
In  last layer, 2 kernels of size $5\times5$ are applied, resulting in a $4\times4\times512$ tensor, which is fed into a sigmoid neuron to obtain final output.

The architecture of the generator is illustrated in Fig. \ref{fig:figure5}.
The generator produces 64-by-64 gray images by taking a 100 dimensional uniform distribution random noises sampled from $[-1,1]$ range as inputs.
The 100 dimensional noise vector is projected and reshaped into a $4\times4\times1024$ tensor with a fully connected layer, and four deconvolutional layers follow this fully connected layer.
The kernel size is $5\times5$, and the stride is $2$ for these layers.
After each deconvolutional layer, the height and width are doubled and the number of channels are halved.
The output sizes of the first three deconvolutional layers are $8\times8\times512$, $16\times16\times256$ and $32\times32\times128$,  respectively.
The last deconvolutional layer outputs a generated $64\times64$ gray image, which is considered as a new sample when the generator is well trained.

The DCGAN model in our DCGAN+L2-SVM framework is trained with $64\times64$ gray images as input. Therefore, in preprocessing stage, all images in  training datasets are resized to $64\times64$ and normalized to $[-1,1]$ range.
After the DCGAN model is well trained, the middle layer activation values (or feature maps) of the discriminator are treated as the deep features of the corresponding input sample, which is fed into the discriminator for further forward propagation .
This procedure is illustrated in Fig~\ref{fig:figure1}.
In the figure, the feature maps are $32\times32\times64$, $16\times16\times128$ and $8\times8\times256$ tensors. These feature maps  are first max-pooled in a  $4\times4$ neighborhood with stride $2$, and the pooled results are $16\times16\times64$,  $8\times8\times128$ and $4\times4\times256$ tensors.
All these deep features are reshaped into a long vector of size $28672$ as a feature representation for the input sample, and then a L2-SVM linear classifier is trained with this type feature representation,  which is used for future identifying whether a pulsar candidate is a real or not pulsar.

\begin{figure*}
	\begin{center}
		\includegraphics[width=0.9\linewidth]{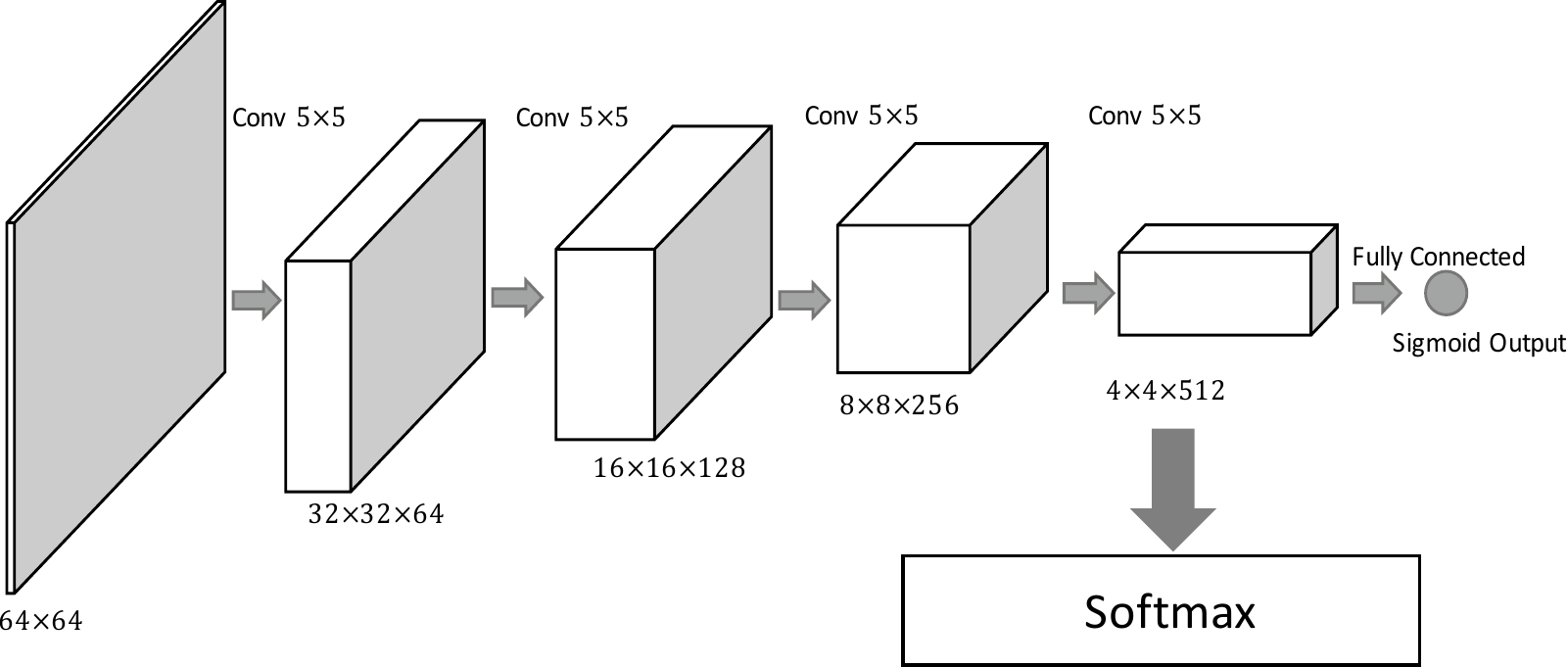}
	\end{center}
	\caption{An illustration of observing  discriminative capability of deep feature representation improvement.}
	\label{fig:figureCapab}
\end{figure*}

\subsection{Training  Strategy for DCGAN}
\label{Sec:DCGANAPS2}

The DCGAN training procedure is a two-player competing game.
The discriminative network aims to predict the correct label of the input sample which is either from the  real training data  data fabricated by the generative model.
The generative model aims to produce synthesized samples that cannot be correctly classified by the discriminative model.
When the training is accomplished, generative network synthesizes samples that are remarkably similar to real pulsar images, and the discriminative network predicts the possibility of the input samples coming from real dataset.

All the models can be trained using the mini-batch stochastic gradient descent (SGD) with a mini-batch size of $M$. To avoid overfitting, the discriminator is optimized in  $k$  steps, and then the generator is updated by one step. Because the activation of the generator output layer is Tanh, the input images are scaled to   $[-1, 1]$, which is the range of the Tanh activation function.
More details about the description of the  training algorithm can be found in \citet{goodfellow2014generative}.

For a better analysis of the improvement of the deep feature representation improvement, the  discriminator in the DCGAN is checked during the DCGAN training process by extracting  deep features using current discriminator for pulsar classification.
 This observation is illustrated in Fig~\ref{fig:figureCapab}.
At a checking point, the  outputs of the last convolutional layer  are fed into a two-class Softmax classifier, and only the SoftMax classifier is updated using the training pulsar samples to minimize the training errors; while those layers in DCGAN stay unchanged  when detecting a change in the discriminative capability. Subsequently, 
the Softmax classifier is used to predict the testing pulsar candidate sample labels, and 
the generalization performance of a classifier is used to judge the  discriminative capability of the deep representation.A higher classification accuracy corresponds to a stronger discriminative capability of the classifier with the learned features.

\section{Results and Analysis}
\label{Sec:Results}
In this section, we first introduce several preliminaries, such as the dataset information, evaluation metrics, and parameter settings. Next, we present and analyse the results of the experiments performed on the HTRU dataset and the PMPS-26k data. Finally, the advantages and disadvantages of the proposed method are discussed.
\subsection{Datasets and Evaluation Metrics}
\label{Sec:Results1}
Two pulsar candidate datasets are investigated in our experiments: the HTRU medlat and the PMPS-26k.

HTRU medlat dataset\footnote{\url{http://astronomy.swin.edu.au/~vmorello/}} is the first publicly available labelled benchmark dataset released by \citet{morello2014spinn}.

HTRU medlat precisely consists of 1,196 positive candidate examples which are sampled from 521 distinct real pulsars, and 89,996 negative candidate examples.
PMPS-26k is an artificially identified pulsar candidate dataset obtained from reprocessing all the PMPS \citep{manchester2001parkes} observation data within the FAST sky (above Declination -14d), which contains 2,000 positive examples that are obtained from real pulsars, 2,000 negative samples that are obtained from non-pulsar signals, 20,000 negative samples that are obtained from radio frequency interference (RFI), and 2,000 samples whose labels are unknown. Such categorization and the associated entries in the database serve as a foundation for the future training of machine-learning algorithms and the pipeline certification to test the validity and efficiency of such a reprocessing and search procedure. 
Table.\ref{tab:table0} lists the number of examples in both the datasets.

\begin{table}
 \caption{The number of samples in HTRU medlat and PMPS-26k datasets. Notably, there are 2,000 samples without labels in PMPS-26k.}	\begin{center}
		\begin{tabular}{|l|c|c|c|}
			\hline
			Dataset Names & \#positive & \#negative  & \#total \\
			\hline\hline
			HTRU medlat & 1,196 & 89,996  & 91,192 \\
			PMPS-26k & 2,000 & 22,000 & 26,000 \\
			\hline
		\end{tabular}
	\end{center}
	\label{tab:table0}
\end{table}

\begin{table}
	\caption{Binary classification confusion matrix. It defines all the outcomes of prediction, which includes True Negative (TN), False Negative (FN), False Positive (FP), and True Positive (TP).}
		\begin{center}
		\begin{tabular}{|l|c|c|}
			\hline
			Outcomes & Prediction - & Prediction + \\
			\hline\hline
			Ground-truth - & True Negative & False Positive \\
			Ground-truth + & False Negative & True Positive \\
			\hline
		\end{tabular}
	\end{center}

	\label{tab:tableC}
\end{table}

The evaluation metrics we adopt for the pulsar candidate classification task here include \textit{Precision}, False Positive Rate (\textit{FPR}), \textit{Recall} and \textit{F-score}.

The binary classification confusion matrix is defined as  in Table. \ref{tab:tableC}, where True Negative (TN, ground-truth and prediction are both negative), False Negative (FN, ground-truth is positive and prediction is negative), False Positive (FP, ground-truth is negative and prediction is positive), and True Positive (TP, ground-truth and prediction are both positive) are defined.

\begin{eqnarray}
Precision &=& \frac{TP}{TP + FP}, \nonumber \\
Recall &=& \frac{TP}{TP + FN},  \nonumber  \\
Fscore &=& \frac{2 * Precision * Recall}{Precision + Recall}.
\end{eqnarray}

\vspace{0.1cm}

Different metrics have a different importance. The false positives cost time, and false negatives cost the scientific discovery of new pulsars, which are relatively rare. From the above equation, we can see that the false positives are correlated with the \textit{Precision}; the \textit{Precision} is higher with a smaller number of false positives. The false negatives are correlated with the \textit{Recall}; the \textit{Recall} is larger with a smaller number of false negatives. Furthermore, the false positive rate \textit{FPR} is defined as follows, which provides the direct rate pertaining to the false positives:

\begin{equation}
FPR = \frac{FP}{ (FP + TN)}
\end{equation}

All of the metrics are in range $[0,1]$. For the \textit{Precision}, \textit{Recall} and \textit{Fscore}, a higher value is more desirable. For the \textit{FPR}, a lower value is more desirable.

\subsection{Experiment Settings}
\label{Sec:Results2}

Two groups of methods are defined in this part: \textbf{Method 1} and \textbf{Method 2}.
The motivation of these two groups is to investigate the impact of different DCGAN models  on the performance.
For the DCGAN, the DCGAN models are trained for the sub-ints and sub-bands, respectively.
However, for the Conditional-DCGAN, one unified DCGAN model is trained for the sub-ints and sub-bands together with a condition vector to identify the sub-ints and sub-bands.

Firstly,  we investigate the method  that is a combination of  DCGAN with L2-SVM (\textbf{Method 1}).
In this method, the parameter settings are similar to those of the DCGAN in \citet{radford2015unsupervised}.
Because the activation function of the generator output is Tanh, the images were scaled to the range of Tanh, $[-1,1]$.
The images in both PMPS-26k and HTRU medlat were resized to 64-by-64 pixels. 
All the weights were initialized by a normal distribution with a mean value of zero and standard deviation of 0.02.
The mini-batch size of SGD was $M=128$, and 
the LeakyReLU adopted 0.2 as the slope of the leak.
The ADAM optimizer \citep{kingma2014adam} was used for accelerating the training, in which the learning rate was set to 0.0002.

The labelled data were first randomly split into three folds:   training, validation and testing with a proportion of 30\%, 30\% and 40\%, respectively.
In this experiment part, three types of classification method were validated and compared: radial basis function kernel SVM (RBF-SVM), CNN and our proposed DCGAN+L2-SVM.
RBF-SVM takes 24 principal components obtained from the principal component analysis (PCA) of input images as the features, and these principal components are adopted to train a RBF-SVM classifier.
While in the experiments, CNN denotes the network architecture used in \citet{zhu2014searching}.   
In their CNN model, the images were firstly resized to $48\times48$ instead of $64\times 64$ pixels.

For our  DCGAN+L2-SVM model, a DCGAN is trained with the positive training samples for the sub-bands and sub-ints, respectively.
The max-pooled activation values of middle layers in the discriminator are regarded as the deep features for each sample.
A L2-loss linear SVM \citep{chang2011libsvm} is then trained with these feature representations. It should be noted  that the parameters in the RBF kernel and SVMs are validated on the validation data to choose the best configuration that leads to  the highest classification accuracy.

For simplicity in later reference, suffixes ‘1’ and ‘2’ refer to the classifiers trained using the time-vs-phase plots and frequency-vs-phase plots, respectively. 

Another method that we investigated war the   Conditional-DCGAN based method (\textbf{Method 2}).
In this method, we partitioned the data into training set and test set.
Only one Conditional-DCGAN  was trained on positive training samples of the sub-bands and sub-ints together.
A condition vector was concatenated the inputs to distinguish sub-bands and sub-ints.

For HTRU medlat data set, two scenarios were investigated.

\textbf{Small Training Set Scenario(STSS)}:
As the number of real positive (RP) examples are limited to 1196, we randomly chose 500 samples for testing and the remaining 696 samples for training. To keep class balance in the training stage, we randomly chose 1196 real negative (RN) samples out of the 89996 for training and testing. Altogether, 1392 examples were considered for the training and testing. In additions, we used the DCGAN to generate positive (GP) samples for training the classifiers. 

\textbf{Large Training Set Scenario(LTSS)}:
The proposed DCGAN model can generate new realistic positive examples to solve the class imbalance problem. We adopted the DCGAN generator to produce 9304 new realistic positive samples. The test set had the same configuration, with 500 real positive and 500 negative samples. As a result, we had 10000 positive examples for training. We also randomly chose 10000 negative examples. Altogether, we had a total of 20000 examples for training. This data partition method is summarized in  Table \ref{tab:table2-1}.

\begin{table}
	\caption{HTRU Medlat Dataset Partition.}
	\begin{center}
		\begin{tabular}{ l | l | l }
			\hline
			Total Sample No.  & Positive (+)   & Negative (-)  \\
			\hline\hline
			Real samples 91192 & 1196 RP & 89996 RN \\
			Small training set 1392  & 696 RP  & 696 RN \\
			Large training set 20000  & 696 RP + 9304  GP  & 10000 RN \\
			The same test set 1000 &  500 RP & 500 RN \\
			\hline
		\end{tabular}
	\end{center}
	
	\label{tab:table2-1}
\end{table}

The second dataset was the PMPS-26k pulsar candidate dataset, which has 2000 positive examples and 20000 RFI examples, as given in Table \ref{tab:table0}. The original examples were directly stored in the form of images with a size of approximately 
$600 \times 800$ pixels.  However, the original sizes were not fixed. We resized all the examples into images with a size of  $ 48 \times 48$ pixels. Each example had a sub-band image and a sub-int image. We partitioned the PMPS-26k dataset into a training set and test set. The training set contained 1500 real positive examples and 1500 real negative examples. The test set contained 500 real positive examples and 500 real negative examples. For a large training set scenario, we generated 8500 realistic positive examples by using the DCGAN, and a total of 10000 positive examples were obtained for the training dataset. Furthermore, in this scenario, 10000 real negative examples were randomly chosen for the testing. The data partition result is summarized in Table \ref{tab:table2-2}.

\begin{table}
	\caption{PMPS-26k Dataset Partition.}
	\begin{center}
		\begin{tabular}{ l | l | l }
			\hline
			Total Sample No.  & Positive (+)   & Negative (-)  \\
			\hline\hline
			Real samples 22000 & 2000 RP & 20000 RN \\
			Small training set 3000  & 1500 RP  & 1500 RN \\
			Large training set 20000  & 1500 RP + 8500  GP  & 10000 RN \\
			The same test set 1000 &  500 RP & 500 RN \\
			\hline
		\end{tabular}
	\end{center}
	
	\label{tab:table2-2}
\end{table}

\subsection{Evaluation Results for DCGAN+L2-SVM (Method 1)}
\label{Sec:Results3}
The performance metrics of the pulsar candidate classification on the HTRU Medlat dataset are presented in Table.\ref{tab:table1}.From this table, we can see that the traditional methods such as the RBF-SVM-1 and RBFSVM-2 achieve an \textit{Fscore} of approximately 0.86. The CNN method has a better performance than that of the RBF-SVMs, with both the baseline methods CNN-1 and CNN-2 demonstrating a satisfactory performance with the \textit{Fscore} values of 0.953 and 0.952, respectively. Specifically, the \textit{Fscore} of CNN-1 is 0.953, which is larger than the corresponding value of 0.868 of the  RBF-SVM-1.   The 
\textit{Fscore} of CNN-2, 0.952 is larger than that of RBF-SVM-2, 0.866.
The reasons for the satisfactory performance of the CNN baselines may be as follows: One reason is that the input data are two-dimensional plots, which convey a large amount of discriminative information for the classification task. The other reason is that the CNNs are able to model the spatial features accurately especially in the case of the two-dimensional images. After training a DCGAN model with a complete training dataset, the learned discriminator in the DCGAN can be used as the hierarchical feature extractor. 
In the framework of DCGAN+L2-SVM,   these hierarchical  features are used to train a L2-SVM linear classifier.
The performance of the DCGAN+L2-SVM is better than that of the baseline CNNs by approximately  $1\%$ in terms of the \textit{Fscore}.
More specifically, DCGAN+L2-SVM-1 exhibits an improvement of  0.6\% over CNN-1, and 
DCGAN+L2-SVM-1 exhibits an improvement of 0.7\% over CNN-2.
These results validate that the DCGAN can provide a reasonable approach for the extraction of deep discriminative features, and the discriminative feature representation can help enhance the pulsar identification accuracy.

\begin{table}
	\caption{Evaluation of different methods (\textbf{Method 1}) on \textit{HTRU Medlat Dataset}. Three types of classifiers were investigated: RBF-SVM, CNN and DCGAN+L2-SVM. Suffix `1' means the model is trained with time-vs-phase plot data, and `2' means the model is trained with frequency-vs-phase plots.}
		\begin{center}
		\begin{tabular}{|l|c|c|c|}
			\hline
			Method & Fscore & Recall & Precision \\
			\hline\hline
			RBF-SVM-1 & 0.868 & 0.854 & 0.882 \\
			RBF-SVM-2 & 0.866 & 0.847 & 0.886 \\
			CNN-1 & 0.953 & 0.956 & 0.950 \\
			CNN-2 & 0.952 & 0.953 & 0.951 \\			
			DCGAN-L2-SVM-1 & 0.963 & \textbf{0.966} & 0.961 \\
			DCGAN-L2-SVM-2 & \textbf{0.964} & 0.963 & \textbf{0.965} \\
			\hline
		\end{tabular}
	\end{center}

	\label{tab:table1}
\end{table}

The performance of the pulsar candidate classification task on the PMPS-26k dataset is presented in Table.\ref{tab:table2}.
It can be seen that the \textit{Fscore}s of RBF-SVM-1 and RBF-SVM-2 are 0.820 and 0.813, respectively.
From the table we can also  see that CNN-1 achieves an  \textit{Fscore} of approximately 0.883, which is 6.3\% larger than that of RBF-SVM-1, and
CNN-2 achieves an \textit{Fscore} of approximately 0.879 , which is 6.6\% larger than that of RBF-SVM-2.
This fact indicates that the CNN methods perform better than the RBF-SVM classifiers on the pulsar candidate classification task.
While DCGAN+L2-SVM-1exhibits an accuracy improvement of approximately 0.6\% over CNN-1 in terms of the \textit{Fscore}, and 
DCGAN+L2-SVM-2 improves accuracy about 0.7\% over CNN-2 on \textit{Fscore}.
The main improvement of DCGAN+L2-SVM over CNN here is in terms of the \textit{recall} value.
Both approaches have an \textit{Fscore} of approximately  0.8.8  
DCGAN+L2-SVM-1 exhibits an improvement of approximately  0.9\% over CNN-1 in terms of  \textit{recall} value, and 
DCGAN+L2-SVM-2 exhibits an improvement of approximately  1.2\% over CNN-2 in terms of  \textit{recall} value.
These results show that the deep feature representations obtained using the DCGAN discriminator output can considerably improve the pulsar candidate classification performance.

To consider  the running time of DCGAN+L2-SVM and CNN, we can analysis the multiply-accumulate operation (MACC) times when network forward inference  is performed in the testing stage. The running time is approximately proportional to MACC. For CNN model of \citet{zhu2014searching}, the MACC is approximately  6.7M\footnote{an on-line tool for estimating MACC: \url{https://dgschwend.github.io/netscope/quickstart.html}}. For our DCGAN discriminator in \textbf{Method 1}, the MACC is approximately 10.41G. For our Conditional-DCGAN discriminator in \textbf{Method 2}, the MACC is approximately  26.28M. We can see that the DCGAN-L2-SVM approach is effective but time-consuming. However, our conditional-DCGAN is considerably faster. With the advancement of the hardware, especially the rapid development of the GPU cards, the float-point calculation is dramatically accelerated.
We suggest the use of a Nvidia Titan XP GPU or higher to perform the testing because a more advanced hardware setting can accelerate the run-time of testing. In addition, many methods of neural network compression and quantization have been proposed in recent years \citep{He2018ADC, Cheng1710Survey}, which makes it possible to simplify the network and decrease the run-time while maintaining a satisfactory performance.

\begin{table}
	\caption{Evaluations of different methods (\textbf{Method 1}) on \textit{PMPS-26k Dataset}.  Three types of classifiers were investigated: RBF-SVM, CNN and DCGAN-L2-SVM. Suffix `1' means the model is trained on time-vs-phase plot data. `2' means the model is learned on frequency-vs-phase plots.}
		\begin{center}
		\begin{tabular}{|l|c|c|c|}
			\hline
			Method & Fscore & Recall & Precision \\
			\hline\hline
			RBF-SVM-1 & 0.820 & 0.806 & 0.835 \\
			RBF-SVM-2 & 0.813 & 0.798 & 0.828 \\
			CNN-1 & 0.883 & 0.886 & 0.881 \\
			CNN-2 & 0.879 & 0.879 & 0.880 \\			
			DCGAN-L2-SVM-1 & \textbf{0.889} & \textbf{0.895} & \textbf{0.885} \\
			DCGAN-L2-SVM-2 & 0.886 & 0.891 & 0.881 \\
			\hline
		\end{tabular}
	\end{center}

	\label{tab:table2}
\end{table}

\subsection{Evaluation Results for Conditional-DCGAN based (Method 2)}
\label{Sec:Results4}

\begin{figure*}
	\begin{center}
		\includegraphics[width=0.9\linewidth]{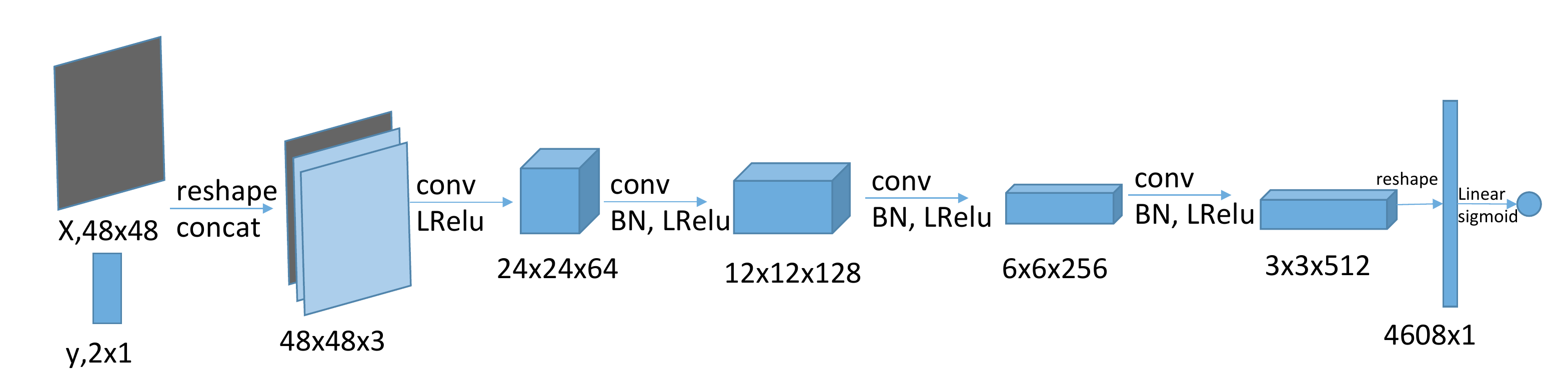}
	\end{center}
	\caption{DCGAN discriminator of  Method 2 for DeepFeature Learning.}
	\label{fig:figure6}
\end{figure*}

\begin{figure*}
	\begin{center}
		\includegraphics[width=0.9\linewidth]{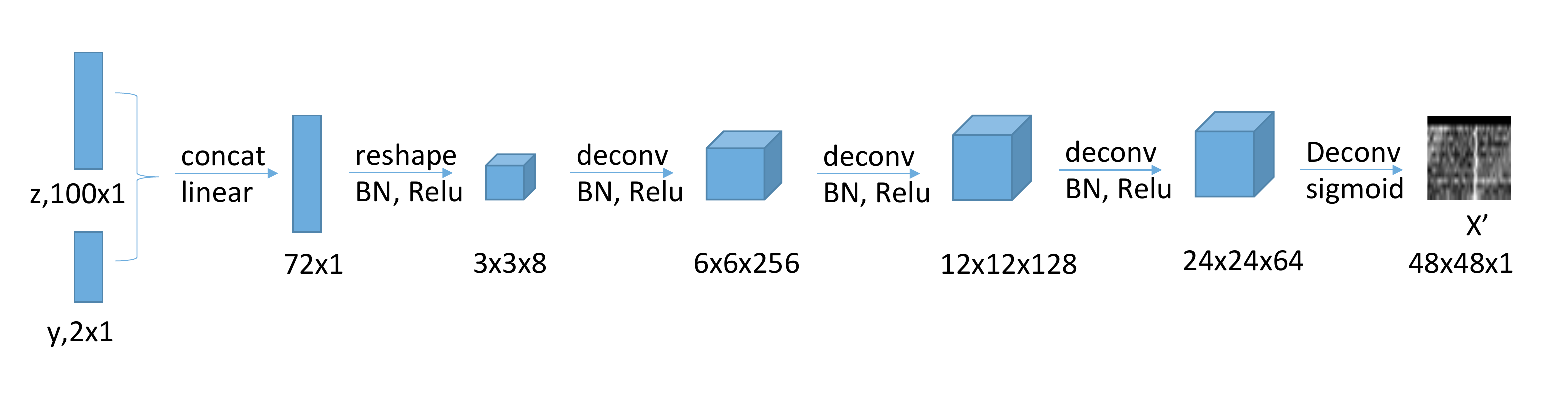}
	\end{center}
	\caption{DCGAN generator of Method 2 for positive sample generation. }
	\label{fig:figure7}
\end{figure*}

In this section, we present the results that using the DCGAN to generate only positive training examples to construct training dataset. Using this generated dataset, several classifiers, including PCA+SVM, DeepFeature+SVM, and CNN have been compared.  Because this framework is different from the previous DCGAN+L2-SVM framework, we call the DeepFeature+SVM an advanced method. The frameworks of the discriminator and the generator are shown in Fig. \ref{fig:figure6} and Fig. \ref{fig:figure7},  respectively. In contrast from the method involving the DCGAN architecture, a conditional-DCGAN is adopted in this case.
A condition feature $y$ is concatenated to inputs, specifically $y=[1,0]$ for sub-bands and $y=[0,1]$ for sub-ints.


For \textit{PCA+SVM}  classifiers, we use 24 principal components as features in training which are also used in Ref. \citep{zhu2014searching}. The PCA transform matrix is calculated using all the samples in dataset, including the real positive,  negative  and DCGAN generated samples.  In Small Training Set Scenario (STSS),  SVMs are trained with regularization  parameter $C=0.05$ for sub-bands, and $C=0.08$ for sub-ints, therefore, the \textit{FPR} can be controlled to be  less than 1\%.  In Large Training Set Scenario (LTSS),  linear SVMs are trained with $C=3000$ for both sub-bands and sub-ints images. 

When training the \textit{DeepFeature+SVM} (DeepF+SVM) classifiers,  we adopt the last convolution layer activation outputs of the DCGAN discriminator as the deep feature representation, that is,  a $4608\times 1$ vector is employed.  In the STSS, we have positive examples in size $696 \times 4608$, and negative examples in size $696 \times 4608$. Parameter $C=0.05$ for sub-bands and $C=0.08$ for sub-ints. In LTSS, we have positive examples in size $10000 \times 4608$ and negative examples in size $10000\times 4608$.  $C=1000$ for both sub-bands and sub-ints images. 

From the SVM classification performance reported in Table \ref{tab:table3} and \ref{tab:table4}, we can see that the LTSS achieves a better performance compared to that of the STSS. This phenomenon indicates that the realistic positive examples generated by the proposed DCGAN are helpful in improving the performance of the pulsar candidate classification. We can also see that the \textit{DeepFeature+SVM} classifiers perform considerably better than the corresponding \textit{PCA+SVM} classifiers. This aspect indicates that our deep features, which are presented as the last convolution layer activation outputs of the DCGAN discriminator are highly discriminative for the classification in the HTRU dataset.

When training the CNN classifiers, we follow the same CNN classifier architecture as used in Ref.\citep{zhu2014searching}. In our experiments, we pre-process the input images and scale the inputs to a range of [-1, 1]. The ground-truth label is 0 or 1. The convolutional layer conv1 has 20 filters of size  $16\times 16$, and its stride is 1. The activation function after the convolutional layers is ({\em Tanh}). The max-pool-1 layer has a filter size of $3\times 3$ . Conv2 has 50 filters sized$8\times 8$, and the stride is 1. Max-pool-2 has a filter size of $2\times 2$ . The final output is a sigmoid activation in the range [0, 1]. The batch size is set as 100, and the learning rate is 0.0002. These two parameters influence the speed of the training convergence. The convergence takes approximately 1 h for one training task of a CNN classifier on one single Nvidia Titan XP GPU card when using the TensorFlow (version 1.0.0) framework on the Python 2.7 platform. The training task occupies 5 GB GPU memory during training.

In the testing stage, the prediction value of the CNN classifier is a sigmoid output which is in range [0,1]. Therefore, we need one threshold value, namely, \textit{thres}, to decide whether a sample is positive or negative. Intuitively, thres controls a trade-off between the recall and FPR. As the FPR is required to be less than 1\%, we should seek a \textit{thres} value that helps decrease the FPR to less than 1\%. In the experiment results, if no \textit{thres} value exists that can reduce the FPR to below 1\%, we choose the \textit{thres} value that makes the classifier have as low an FPR value as possible.

We also evaluated these classifiers with the PMPS-26k dataset. Because the hyperparameter in the classifiers is data dependent, for the \textit{PCA+SVM} classifiers, we set  $C=10^{-4}$ for sub-bands,  and $C=10^{-1}$  for the sub-int images in the small training set scenario. By adjusting the $C$ values in the SVM training, the minimum FPR for the subband \textit{PCA+SVM} classifier is 0.33. The minimum FPR for the sub-int \textit{PCA+SVM} classifier is 0.25.

For \textit{DeepFeature+SVM} classifiers,  $C=10^{-2}$ for small training set scenario both in sub-bands and sub-ints images. $C=10$ for large training set scenario both in sub-bands and sub-ints. The evaluation results are shown in Table \ref{tab:table4}.

From the results presented in Table  \ref{tab:table3} and \ref{tab:table4}, we can see that:
\begin{enumerate}
\item The DCGAN generated positive examples are remarkably helpful in improving the pulsar classification performance in the \textit{PCA+SVM} classifiers.
\item The deep features obtained from the last convolutional activations of the DCGAN discriminator are highly discriminative in the pulsar classification.
\item The \textit{CNN} classifiers perform better than the  \textit{PCA+SVM} classifiers according to the recall value, but they are not as effective as the SVM+deep-DCGAN features. Furthermore, the DCGAN generated positive examples are not as helpful for the \textit{CNN} classifiers. The \textit{CNN} classifiers also have a larger FPR on the real data ($1.8$\% for the sub-bands and $2.2$\% for the sub-ints, as shown in Table \ref{tab:table3}) under the small training set scenario, which limits their value in practical use.
\end{enumerate}

\begin{table}
	\caption{Recall and FPR Evaluations of different methods (\textbf{Method 2}) on \textit{HTRU Medlat Dataset}.  }
	\begin{center}
		\begin{tabular}{|l| c | c|c| c | }
			\hline
			Classifier \& dataset &  Recall & Precision & F1-score & FPR  \\
			\hline
			\textit{PCA+SVM}& & &&\\
			   sub-bands(STSS) &  {\bf 0.654} & 0.827 & 0.730& {\bf 0.0}\\
			   sub-bands(LTSS) &  {\bf 0.744} & 0.872 & 0.803& {\bf 0.0}\\
		            sub-ints(STSS) &  {\bf 0.612} & 0.804 & 0.695& {\bf 0.004}\\
			   sub-ints(LTSS) &  {\bf 0.620} & 0.808 & 0.702& {\bf 0.004}\\		   
			\hline 
			\textit{DeepF+SVM}& & &&\\
			   sub-bands(STSS) & {\bf 1.0} & 1.0 & 1.0 & {\bf 0.0}\\
			   sub-bands(LTSS) &  {\bf 1.0} & 1.0 & 1.0  & {\bf 0.0}\\
		            sub-ints(STSS) &  {\bf 1.0} & 1.0 & 1.0 & {\bf 0.0}\\
			   sub-ints(LTSS) &  {\bf 1.0} & 1.0 & 1.0& {\bf 0.0}\\		   
			\hline
                             \textit{CNN}& & &&\\
			   sub-bands(STSS) &  {\bf 0.928} & 0.955 & 0.941& { 0.018}\\
			   sub-bands(LTSS) &  {\bf 0.818} & 0.907 & 0.860& { 0.004}\\
		            sub-ints(STSS) &  {\bf 0.878} & 0.928 & 0.902& {0.022}\\
			   sub-ints(LTSS) &  {\bf 0.754} & 0.874 & 0.810& { 0.006}\\	
			\hline
		\end{tabular}
	\end{center}
	\label{tab:table3}
\end{table}

\begin{table}
	\caption{Recall and FPR Evaluations of different methods (\textbf{Method 2}) on \textit{PMPS-26k Dataset}.  }
		\begin{center}
		\begin{tabular}{|l| c | c|c| c | }
			\hline
			Classifier \& dataset &  Recall & Precision & F1-score & FPR  \\
			\hline
			\textit{PCA+SVM}& & &&\\
			   sub-bands(STSS) &  {\bf 0.814} & 0.742 & 0.776& {\bf 0.330}\\
			   sub-bands(LTSS) &  {\bf 0.840} & 0.734 & 0.783& {\bf 0.372}\\
		            sub-ints(STSS) &  {\bf 0.730} & 0.740 & 0.735& {\bf 0.250}\\
			   sub-ints(LTSS) &  {\bf 0.710} & 0.731 & 0.720& {\bf 0.248}\\		   
			\hline 
			\textit{DeepF+SVM}& & &&\\
			   sub-bands(STSS) & {\bf 1.0} & 0.999 & 0.9995 & {\bf 0.002}\\
			   sub-bands(LTSS) &  {\bf 0.998} & 0.999 & 0.998 & {\bf 0.0}\\
		            sub-ints(STSS) &  {\bf 1.0} & 1.0 & 1.0 & {\bf 0.0}\\
			   sub-ints(LTSS) &  {\bf 1.0} & 1.0 & 1.0& {\bf 0.0}\\		   
			\hline
                             \textit{CNN}& & &&\\
			   sub-bands(STSS) &  {\bf 0.904} & 0.890 & 0.897& { 0.124}\\
			   sub-bands(LTSS) &  {\bf 0.806} & 0.902 & 0.851& { 0.002}\\
		            sub-ints(STSS) &  {\bf 0.880} & 0.916 & 0.898& {0.048} \\
			   sub-ints(LTSS) &  {\bf 0.858} & 0.921 & 0.888 & { 0.016}\\	
			\hline
		\end{tabular}
	\end{center}
	\label{tab:table4}
\end{table}

On the same dataset, HTRU Medlat Dataset, we compare the three algorithms in terms of the recall and FPR also.  The results are shown in Table \ref{tab:table5}.

\begin{table}
	\caption{Comparison of classification performance for different classifiers on \textit{HTRU Medlat Dataset}.  }
		\begin{center}
		\begin{tabular}{|l| r | r|r| }
			\hline
			Classifier  &  Recall  & FPR  \\
			\hline
			\citet{morello2014spinn}	&  100 \% & 0.64 \% \\
			  \citet{lyon2016fifty}    &  92.8 \% & 0.5 \% \\
		             \citet{yaoyao2016cis}  &   97.49 \%  & 0.0 \\
			   DeepF+SVM &  100 \% & 0.0\\
			   \hline		   
			 \end{tabular}
	\end{center}
	\label{tab:table5}
\end{table}

As we know, \citet{morello2014spinn}  carefully designed 5 types  of features and fed them into a shallow neural network.  They achieved a 100\% recall rate 
 when the false positive rate was 0.64\% on HTRU dataset. In our experiments, we trained DCGAN model on the real positive examples, and extracted 4068 activation values of the last convolutional layer as  the deep feature. For the same dataset, we obtained a 100\% recall when false positive rate was 0 on all the \textit{DeepFeature+SVM} classifiers, irrespective of the type of of the input image (either phase-frequency or phase-time patterns).

\subsection{Discriminative Capability Analysis}
\label{Sec:Results5}
\begin{figure}
	\begin{center}
		\includegraphics[width=0.95\linewidth]{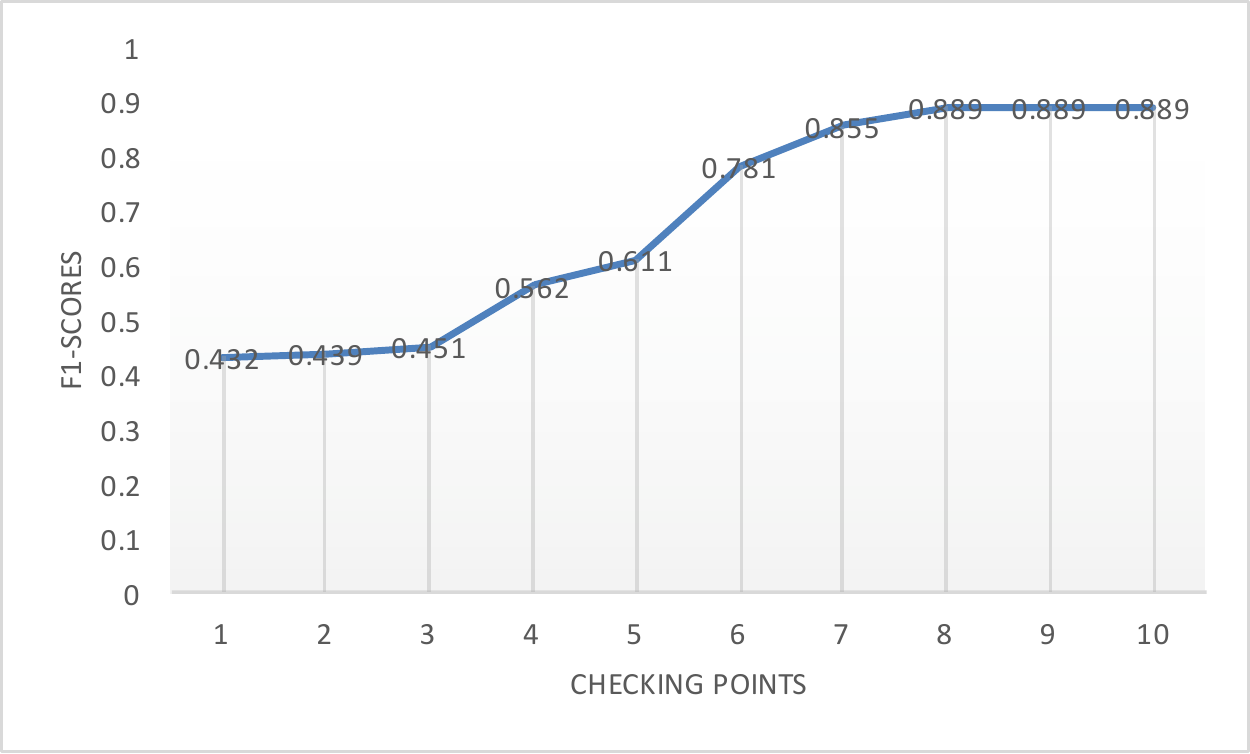}
	\end{center}
	\caption{A performance curve showing  the discriminative capability changes over DCGAN training process.}
	\label{fig:figureCapaChg}
\end{figure}

To analyze the impact  of the learnt deep feature representation on the discriminative capability, we designed a checking technique as shown in Fig~\ref{fig:figureCapab}. In this experimental investigation of the discriminative capability,   PMPS-26k dataset with time-vs-phase plot images was utilized to train the DCGAN.
During the training process of DCGAN on all the  labelled samples, we set a checking point at every 200 epochs of iteration. 
At each checking point, we trained a Softmax classifier by using the training data composed with the deep features and obtained a classification F1-score computed with  the testing data.
Fig~\ref{fig:figureCapaChg} shows the \textit{F-score} results of the pulsar candidate classification at 10 different checking points. The curve illustrates that the \textit{F-score} increases along with the training process and reaches to 0.889 as the DCGAN training error converges to a stationary point.
This result illustrates that the discriminative capability of the DCGAN learnt deep feature representation is enhanced with the increase in the number of epochs in the DCGAN training process.

\subsection{Discussions and consideration of further work }
\label{Sec:Results6}
The experimental results shown in Table \ref{tab:table1} and Table \ref{tab:table2} show that the proposed DCGAN+L2-SVM framework exploits the advantages of the deep convolutional generative adversarial network and achieves a satisfactory performance for the automatic pulsar candidate identification task. Even when the number of training samples, specifically that of the positive ones, is relatively limited, the deep convolutional generative adversarial learning can still provide a reasonable technique to learn a strong discriminative feature representation. The principle behind this framework is that the generator can produce fake samples that are almost the same as the given training samples. For the class imbalance problem, one of the tactics is to sample an equal number of data points for each class from the given dataset. However, for the pulsar candidate dataset, only a very small number of training samples can be drawn because of the limited number of positive pulsar samples; for example, for the HTRU Medlat dataset, only 1,196 training sample pairs exist, and for the PMPS-26k dataset, only 2,000 training sample pairs can we obtained in this manner. It is well known that to train a deep neural network (DNN), a large number of labelled samples are needed. If a small sample dataset is used to train the DNN, we cannot expect to obtain a satisfactory generalization performance because overfitting can occur. This aspect is clarified by comparing the CNN model performance on the HTRU Medlat and PMPS data. To address this problem, we propose the adoption of the tactic of generating synthetic samples. During the training process, the generator in the DCGAN generates fake samples, and the discriminator is trained with true and fake samples. After the training loss converges to a stationary Nash equilibrium, we can obtain not only a capable generator but also a capable discriminator. Using this framework, we can solve the class imbalance and small training sample problem simultaneously. Another advantage of the DCGAN+L2-SVM framework is that it is possible to eliminate the challenging work of designing satisfactory handcrafted features for classification by learning an end-to-end model with the original input data. The results on both the datasets show that the DCGAN+L2-SVM model has the capability of learning stronger discriminative features for pulsar classification than that learned by the traditional CNN model. As a complicated deep neural network model, the DCGAN+L2-SVM can perform better with limited labelled training data under the training strategy of the GAN; in this context, this model outperforms the CNN models. It is also believed that with more new labelled samples, especially positive real pulsar candidate data, the DCGAN+L2-SVM and CNN models can both further improve the pulsar candidate classification performance.

We already know that the number of real pulsar samples is limited. Another important aspect that we must consider is the diversity of real pulsars. The characteristic of the real pulsar class cannot be completely learned if the training real pulsar samples do not exhibit a reasonable diversity. For example, if certain unique real pulsars are missed, our learned DCGAN model cannot generate similar unique training samples. Furthermore, the final pulsar selection tool would fail to recognize the unique pulsars that the tool has never encountered previously. This aspect might be one shortcoming of machine-learning-based methods. Therefore, for unique pulsars, more training data are required or human experts may be considered to address such candidates.

We also proposed a new conditional-DCGAN based method to generate both sub-int and sub-band realistic positive samples together. This aspect is achieved by concatenating a condition vector $y$ to the inputs, specifically,$y=[1,0]$ for the sub-bands and $y=[0,1]$ for the sub-ints. While generating only positive examples, the conditional DCGAN generator outputs the corresponding results according to the condition vector. The evaluations of the DeepFeature+SVM classifiers are surprisingly satisfactory, as shown in  Table \ref{tab:table3} and Table \ref{tab:table4}. However, a decrease in the performance can be noted in Table \ref{tab:table4}. The DeepFeature+SVM of the sub-band (LTSS) evaluations exhibits a worse performance than that of the STSS. We argue that the cause is a lack of diversity in the real training positive samples. As is known, most available pulsars have only a small number of unique ones, and wide or multi-peak pulse profiles or harmonic signals are often rare. The trained conditional-DCGAN learns the distribution of only the limited real positive samples, which may contain only a small number of unique pulsars; in such a case, the generated realistic positive samples lack in diversity. As a result, the performance of the APCI system is degraded, and this case is more obvious for the CNN classifiers in both the tables. Therefore, improving the diversity of the training data could be helpful to achieve a better APCI system performance.

From the results shown in Table \ref{tab:table5}, it can be concluded that the deep features are extracted as the activation values of the last convolutional layer in the DCGAN discriminator. This feature representation is highly discriminative and remarkably helpful in the classification of pulsar candidates. Furthermore, in our \textit{DeepFeature+SVM} classifier training, only one type of input image is used (either sub-band or sub-int). However, the other three methods take a group of input signals, including sub-bands, sub-ints, DM-SNR curves and other signals, to extract the features. 

The AI-based methods including the CNN and our proposed DCGAN-L2-SVM and conditional-DCGAN exhibit better results on the HTRU has better results according to the metrics. Only for the PCA+SVM method are the metrics of the PMPS-26k better than the corresponding values for the HTRU. There could be several reasons for the differences in the performance on the two datasets. One reason could be that the training data quality is different for the two datasets. Another reason may be that the samples in one dataset are more difficult to recognize. For example, the RFI-produced candidates may be too similar to the real pulsars, or the real pulsars may contain excessively many rare types, which the AI-based models may not be well trained for. Although there exists a performance difference on the different datasets, re-designing a more customized network architecture for a specific dataset by performing a network architecture search \citep{zoph2018learning, ZophL16google}  could help improve the results. 

In this work, we mainly evaluated the performance of the methods based on deep neural networks, and only two-dimensional diagnostic plots were taken as the network inputs. In later work, we will design a multi-modal input DNN based pulsar recognition model, which will incorporate one-dimensional SPH and DM curves, two-dimensional TPP and FPP, and other hand-crafted features together to make the pulsar identification more accurate. Certain traditional machine learning models that use hand-crafted features of the SPH or DM curve to train the classifiers provide complementary discriminating capabilities. For example, we built SVM-based ensemble nets to recognize the pulsar candidates and attained a higher classification accuracy \citep{yaoyao2016cis}. We will investigate these one-dimensional inputs with a stacked auto-encoder deep neural network architecture and develop the fast learning algorithm \citep{guo2004pseudoinverse,wang2017new, Guo2017pilae}  for the stream data analysis. Because ensemble neural networks can also improve the pulsar candidate classification abilities, the assembly of the deep models and traditional models together to attain a better pulsar identification performance will be studied in our later work. 

In our future work, we will consider building an artificial intelligence system with various techniques \citep{Guo2018ijcis} to address the pulsar data processing problem to reinvent the study of the pulsar search procedure, especially to reduce the DM parameter search processing time, and apply the DCGAN+L2-SVM model to change the traditional pipeline of pulsar searching by taking the raw data as the input for classification before the candidates are folded.  

\section{Conclusion}
\label{Sec:Conclusion}

In this work, we proposed a DCGAN-based automatic pulsar candidate identification framework called DCGAN+L2-SVM. In the DCGAN+L2-SVM framework, a DCGAN model \citep{radford2015unsupervised} is trained on all the labelled balance class samples, and the max-pooled middle layer activation values of the discriminator in the DCGAN are regarded as the hierarchal feature representation; this stage can be considered as an unsupervised learning stage. When the DCGAN is well trained, these middle layer activation values of the discriminator are taken as the training sample vectors to perform the supervised training of an L2-SVM classifier  \citep{chang2011libsvm}. This trained L2-SVM classifier is used for identifying the new input of a pulsar candidate to be a real pulsar or not. The results of the experiments performed on the HTRU Medlat dataset and our self-collected PMPS-26k dataset show that the proposed framework can learn strong discriminative features for the pulsar candidate selection task, and the DCGAN+L2-SVM outperforms a previously used CNN as well as the RBF-SVM models.
We also investigated a scenario in which the conditional DCGAN based method is adopted for generating realistic positive examples. The evaluations demonstrated the value of both the DCGAN deep features and the generated positive samples for the pulsar identification.

\section*{Acknowledgements}
This work was fully supported by grants from  the Joint Research Fund
in Astronomy (U1531242) under cooperative agreement between
the National Natural Science Foundation of China (NSFC) and the
Chinese Academy of Sciences (CAS), and the National Natural Science Foundation of China (61375045, 11690024, 11603046), and supported by the CAS Strategic Priority Research Program No. XDB23000000.
\bsp	
\label{lastpage}
\end{document}